\documentclass[
    aps,
    prx,
    letterpaper,
    nobalancelastpage,
    twocolumn,
    superscriptaddress,
    nofootinbib,
    longbibliography
]{revtex4-2}

\usepackage[colorlinks, citecolor=blue]{hyperref}
\usepackage{graphicx}
\usepackage{amsmath}
\usepackage{amssymb}
\usepackage[english]{babel}
\usepackage{color}
\usepackage[version=4]{mhchem}
\usepackage{dsfont}
\usepackage{multirow}
\usepackage{booktabs}
\usepackage{array}
\usepackage{physics}
\usepackage{bm}
\usepackage{comment}
\usepackage{braket}
\usepackage{slashed}
\usepackage{float}

\usepackage{soul}

\usepackage{textcomp}

\graphicspath{{./Figures/}}

\renewcommand{\t}[1]{\text{#1}}
\newcommand{\termsym}[3]{{}^{#1}\text{#2}_{#3}}
\newcommand{\p}{\partial}
\newcommand{\deldel}[2]{\frac{\delta #1}{\delta #2}}

\newcommand{\UIUC}{
    Department of Physics,
    The University of Illinois at Urbana-Champaign,
    Urbana, IL 61801, USA
}

\renewcommand{\cite}[1]{\mbox{\citep{#1}}}

\addto\captionsenglish{}
\addto\captionsenglish{\renewcommand\thefigure{\arabic{figure}}}

\newcommand{\fidemswap}{0.9925}
\newcommand{\fidemcz}{0.9929}
\newcommand{\fidenmccz}{0.9814}
\newcommand{\fidemshelve}{0.9907}

\begin{document}

\title{Three-qubit encoding in ytterbium-171 atoms for simulating 1+1D QCD}
\author{William Huie}\altaffiliation{These authors contributed equally to this work}
\affiliation{\UIUC}
\author{Cianan Conefrey-Shinozaki}\altaffiliation{These authors contributed equally to this work}
\affiliation{\UIUC}
\author{Zhubing Jia}
\affiliation{\UIUC}
\author{Patrick Draper}\email{pdraper@illinois.edu}
\affiliation{\UIUC}
\author{Jacob P. Covey}\email{jcovey@illinois.edu}
\affiliation{\UIUC}

\begin{abstract}
Simulating nuclear matter described by quantum chromodynamics using quantum computers is notoriously inefficient because of the assortment of quark degrees of freedom such as matter/antimatter, flavor, color, and spin. Here, we propose to address this resource efficiency challenge by encoding three qubits within individual ytterbium-171 atoms of a neutral atom quantum processor. The three qubits are encoded in three distinct sectors: an electronic ``clock" transition, the spin-1/2 nucleus, and the lowest two motional states in one radial direction of the harmonic trapping potential. We develop a family of composite sideband pulses and demonstrate a universal gate set and readout protocol for this three-qubit system. We then apply it to single-flavor quantum chromodynamics in 1+1D axial gauge for which the three qubits directly represent the occupancy of quarks in the three colors. We show that two atoms are sufficient to simulate both vacuum persistence oscillations and string breaking. We consider resource requirements and connections to error detection/correction. Our work is a step towards resource-efficient digital simulation of nuclear matter and opens new opportunities for versatile qubit encoding in neutral atom quantum processors.
\end{abstract}
\maketitle

\section{Introduction}
\label{Intro}
In the era of noisy intermediate-scale quantum devices~\cite{Preskill2018}, resource efficiency is crucial for utilizing quantum hardware in the short term. As many quantum hardware platforms begin to hit scaling barriers, it is increasingly important to optimize the use of resources such as qubit count, connectivity, and coherence time. A paradigmatic example of this need is the digital quantum simulation of nuclear physics phenomena~\cite{Banuls2020,Paulson2021,Farrell2023,Farrell2023b}. Simulations of quark matter must contend with several degrees of freedom including flavor, color, matter/antimatter, and spin; and they must also address the underlying fermionic statistics which require the use of notoriously inefficient fermion-to-qubit mapping schemes such as the Jordan-Wigner transformation~\cite{Jordan1928,Bravyi2002,Gonzalez2023}. Additionally, gauge bosons that mediate interactions between quarks add the challenge of encoding bosonic state spaces that could be large~\cite{Crane2024}.

\begin{figure}[t!]
    \includegraphics[width=\linewidth]{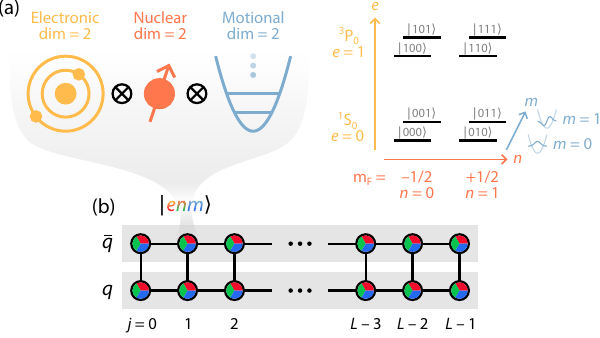}
	\caption{
        \textbf{Overview of the platform}.
        (a) We encode individual qubits in each of the electronic $\termsym{1}{S}{0}$--$\termsym{3}{P}{0}$ (yellow, `$e$'), nuclear Zeeman (orange, `$n$'), and two lowest-energy states of the motional (blue, `$m$') degrees of freedom. The tensor product of these three two-dimensional Hilbert spaces forms the computational `quoct' space. (b) We consider two parallel chains of quocts to encode states for the purpose of a lattice QCD simulation in 1+1D. We consider two parallel 1D chains of quocts, each comprising $L$ lattice sites. At each site, two quocts are used to represent the presence or absence of quarks ($q$) or anti-quarks ($\bar{q}$) in a simulated space of three colors of a single flavor. The $e$, $n$, and $m$ qubits of each quoct encode the presence of a red, green, and blue (anti-)quark at each lattice site.
    }
    \label{fig:overview}
\end{figure}

A natural solution to this challenge is to use larger code spaces in the underlying quantum information carriers of a given hardware platform~\cite{Farrell2023, Farrell2023b}. Specifically, instead of quantum bits (qubits) with a $d=2$ code space, larger-$d$ quantum spaces called qudits could be employed to pack more information into the same physical entity. In superconducting circuits, this could be accomplished with bosonic encodings~\cite{Terhal2020, Cai2021, Crane2024} or with higher-lying states in anharmonic systems like transmons~\cite{Fischer2023, Wang2025}. In trapped ions and neutral atoms, the most natural approach to increasing the code space is to use more internal Zeeman states. Indeed, large code spaces are actively being pursued with nuclear and electronic spins, often combined with long-lived metastable states~\cite{Low2020, Omanakuttan2021, Ringbauer2022, Hrmo2023, Omanakuttan2023, Low2023, Omanakuttan2024, Zalivako2024}. Motional modes of the atoms' harmonic trapping potential offer another option~\cite{Jost2009, Fluhmann2019, Scholl2023b, Grochowski2023, Crane2024, Liu2024a, Liu2024b, Bohnmann2025}. Code spaces of $d=3$ and $d=5$ have recently been exploited to efficiently represent gauge bosons in quantum electrodynamics~\cite{Meth2023}.

However, a major challenge for qudits is the establishment of universal gate sets, the development of protocols for benchmarking and tomography, and the transpilation of algorithms designed for qubit-based hardware. Although gauge bosons may naturally be mapped to a qudit system~\cite{Meth2023}, quark matter for instance shares certain behavior with qubits since fermionic statistics forces binary occupancy. A partial solution is a hybrid qubit-qudit processor in which matter particles are represented with qubits while gauge fields are represented with qudits~\cite{Meth2023}. However, in general, it remains difficult to fully exploit the potential advantage of qudits and to demonstrate their viability as an alternative to qubit-based processors.

Here, we propose an approach that offers the advantages of both qubits and qudits. We focus on a $d=8$ ``quoct" encoded in neutral ytterbium-171 atoms in optical tweezers, but we ``qubitize" this space into three distinct qubits that arise from distinct degrees of freedom in the atom~\cite{Campbell2022,Jia2024}. Specifically, we employ an electronic qubit, $|e\rangle$, based on the optical ``clock" transition; a nuclear spin qubit, $|n\rangle$, based on the spin-1/2 nucleus; and a motional qubit, $|m\rangle$, based on the lowest two harmonic oscillator states of the atom along a radial direction of the tweezer trap (see Fig.~\ref{fig:overview}). We develop a family of composite sideband pulses that isolate the lowest two motional states. We present a qubit-based universal gate set for intra- and inter-quoct operations that include high-fidelity intra-quoct CZ, SWAP, and CCZ gates and inter-quoct C$\bar{\text{C}}$Z gates. 

Further, we showcase the versatility of our three-qubit encoding platform and its utility by developing circuits to simulate single-flavor quantum chromodynamics (QCD) in 1+1D. By choosing the axial gauge, we obviate the need for gauge bosons in one spatial dimension, enabling a direct mapping of the occupancy of the three quark colors onto our three qubits (see Fig.~\ref{fig:overview}). We use a row of atoms to represent quarks and another row to represent antiquarks, and we construct Trotterized circuits that, with only a single site (two atoms), faithfully simulate vacuum persistence and string breaking dynamics. Our work lays the foundation for other multi-qubit encodings and circuit compilations, and it takes a step towards unlocking the potential of quantum hardware for resource-efficient simulation of large-scale nuclear matter. 

\section{The Quoct architecture}
\label{quoct}

Our three-qubit quoct architecture with $^{171}$Yb was inspired by, and is an extension of, our recently-proposed two-qubit ``ququart" architecture based on the electronic optical ``clock" transition and the nuclear spin-1/2 degree of freedom \cite{Jia2024}. One of the main new ingredients here is the addition of the third qubit, which is encoded within the harmonic oscillator states of the atom's motion within the optical trapping potential. Given the recent progress on cooling optically trapped atoms to their motional ground state with high probability \cite{Kaufman2012, Thompson2013, Jenkins2022, Scholl2023b}, encoding quantum information in motional degrees of freedom has become an active new field for neutral atom arrays \cite{Scholl2023b, Grochowski2023, Crane2024, Liu2024a, Liu2024b, Bohnmann2025}. The motional degree of freedom enjoys long coherence times since it is decoupled from nearly all electromagnetic noise. Indeed a coherence time of $T_2^*\approx100$ ms has been observed for atoms in tweezers \cite{Scholl2023b}, which is primarily limited by intensity noise that can be straightforwardly removed. 

We choose to employ a motional qubit in this work instead of, e.g., using a larger nuclear spin manifold because we want three pseudo-independent degrees of freedom in which to encode our three qubits. Hence, the motional sector offers an opportunity to add a qubit without sacrificing the established capabilities with our two-qubit system of electronic and nuclear degrees of freedom \cite{Jia2024}, and we note that it may be possible to add two more motional qubits by leveraging motion in the other two dimensions within the three-dimensional optical potential. The three motional degrees of freedom are sufficiently separate due to the intentional or accidental non-isotropic nature of the optical trap. The mode is selected by momentum conservation; since we couple to the motional states via the momentum of the photon that drives the electronic transition, the alignment of the laser beam with respect to the optical trap determines the projection onto the motional eigenbasis. We assume that the laser couples to a single radial motional mode \cite{Scholl2023b}.

For the remainder of Section \ref{quoct}, we will denote three-qubit quoct states in the form $\ket{enm}$ and two-qubit ququart states in the form $\ket{en}$ (where $e, n, m \in \{0, 1\}$). Further, bipartite states and gates with an explicit tensor product will refer to the quoct or ququart states (or gates thereon) of two adjacent atoms.

\subsection{Intra-quoct gates and motion-selective operations}
\label{intra}

As an extension of the ququart architecture explored in previous work \cite{Jia2024}, the quoct space inherits a significant portion of its supported operations from that architecture with identical implementations. All operations involving the ``clock'' transition ($e$ qubit) and the nuclear spin-1/2 degree of freedom ($n$ qubit) are subsumed under the quoct gate set. For the purposes of this work (see Fig. \ref{fig:circuits}), we specifically inherit the following operations: phase rotation gates to arbitrary angles on the $e$ and $n$ qubits ($Z_e(\alpha)$, $Z_n(\alpha)$) via spin-selective light shifts applied to pairs of ququart states; bit-flip and Hadamard gates on the $e$ qubit ($X_e$, $H_e$) via $\pi$-polarized pulses on the clock transition; bit-flip and Hadamard gates on the $n$ qubit ($X_n$, $H_n$) via simultaneous stimulated Raman pulses within the $e = 0$ and $e = 1$ subspaces mediated by states in the $\termsym{3}{P}{1}$ and $\termsym{3}{D}{1}$ manifolds, respectively; \textit{e}-\textit{n} swap gate ($\text{SWAP}_{e,n}$) via a $\sigma^+$-polarized pulse on the $\ket{01} \leftrightarrow \ket{10}$ clock transition. Here, we also introduce a new ququart operation, which is a $\pi$-pulse on the $e$ qubit conditioned on $n = 1$ ($\text{CNOT}^n_e$). This, as well as $X_e$, $X_n$, and $\text{SWAP}_{e,n}$ are based on a CORPSE (Compensation for Off-Resonance with a Pulse SEquence)-based \cite{Cummins2003} motion-preserving pulse (MPP) \cite{Lis2023} on their respective transitions, while $H_e$ and $H_n$ use the detuned pulses explored in Ref. \cite{Jia2024}. Notably, MPPs only implement a $\pi$ rotation; for rotations through general angles $\alpha$ about the X-axis of a given transition ($X_{\{e,n\}}(\alpha)$), we resort to conjugation of the appropriate $Z_{\{e,n\}}(\alpha)$ gate with $H_{\{e,n\}}$.

\begin{figure*}[t!]
    \includegraphics{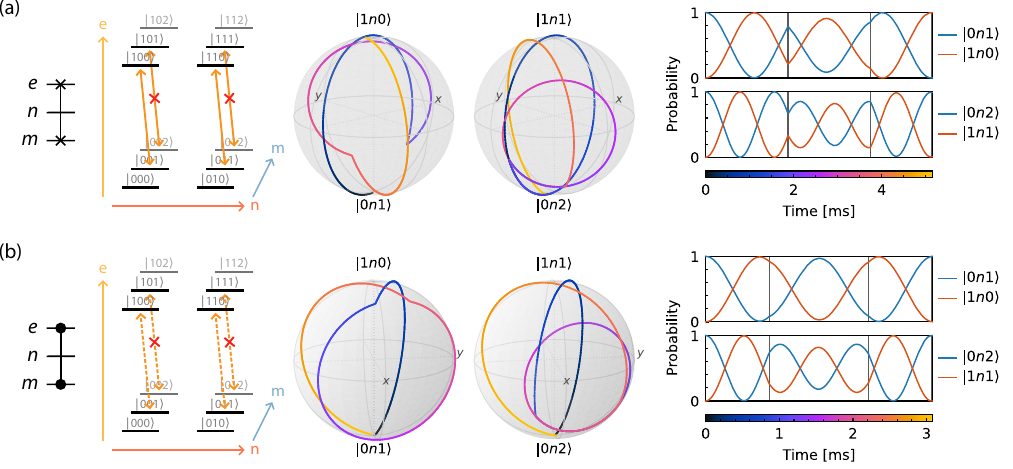}
    \caption{
        \textbf{Motion-selective intra-quoct gate sequences}.
        Pulse sequences showing implementations of the $e$-$m$ SWAP (a) and $e$-$m$ controlled Z (b) gates with fidelities $\mathcal{F}_{\text{SWAP}_{e,m}} \approx \fidemswap$ and $\mathcal{F}_{\text{CZ}_{e,m}} \approx \fidemcz$, respectively. For both gates, driven couplings between the relevant quoct states and next-lowest ($m = 2$) motional states are shown, and couplings for which the pulse sequence performs a $2\pi$ rotation are marked with a red X. Dashed arrows indicate closed state trajectories through which a phase rotation is performed, rather than population transfer. Simulated Bloch sphere trajectories and flattened probability time traces for the relevant pairs of states are also shown. The pulse sequences are agnostic to the $n$ qubit because the driving light pulse is $\pi$-polarized.
    }
    \label{fig:gate-seqs}
\end{figure*}

As with ququarts, operations within the quoct space are dependent on inherited compensatory light shifts required to maintain suitable control over the constituent atomic states. Namely, ququart operations require a passive, ``background'' light shift applied to one of the ququart states to equalize the nuclear Zeeman splittings in the $e = 0$ and $e = 1$ subspaces in order to allow the \textit{n}-preserving clock transitions to be driven simultaneously with a monochromatic pulse. Secondly, the $\text{CNOT}^e_n$, $\text{CNOT}^n_e$, and $\text{SWAP}_{e,n}$ gates both require compensatory light shifts to correct phases accrued by pairs of ququart states during each operation. In the first case, phase is accrued by the $\ket{10}$ and $\ket{11}$ states due to the operation itself; in the latter two cases, phases are accrued by $\ket{00}$ and $\ket{10}$, and $\ket{00}$ and $\ket{11}$ due to the proximity of the clock transition's wavelength to those of other transitions accessible from the $e = 0$ subspace, primarily to $\termsym{3}{P}{1}$ states. In all cases, accrued phases are corrected with additional light shifts applied to individual ququart states using various other nearby transitions, e.g. those from $e = 0$ states to $\termsym{3}{P}{1}$ or from $e = 1$ states to $\termsym{3}{D}{1}$ \cite{Li2025}.

When extending the computational space to include motional modes, care must be taken to ensure that no ququart operations disturb information encoded in the additional degree of freedom. As specified above, we restrict to only the bottom two motional states and, by construction, MPP-based operations preserve the motional ground state with high fidelity. We find through simulation that this naturally extends to superpositions of the targeted motional states as well, up to a small relative phase ($\lesssim 1^\circ$) between them and a global phase on all states undergoing dynamics due to the MPP. Up to this global phase, we find that MPPs operate on general quoct states with fidelity $\gtrsim 0.999$. Similarly, the Hadamard gate implementation designed in Ref. \cite{Jia2024} preserves superpositions of motional states up to a $\approx \pi$ relative phase within the motional sector with fidelity $\gtrsim 0.999$. We examine MPPs and the Hadamard gates in more detail in Appendix \ref{mpp-analysis}. Any erroneous phases can be corrected after the MPP operation using either selectively applied light shifts within the ququart space or a $Z_m$ gate on the motional qubit (described below). On the other hand, the inherited phase operations ($Z_e$ and $Z_n$, as well as state-selective light shifts) can all be made entirely agnostic to motion by using light fields that are locally flat in intensity over the length scale of the targeted motional states.

\begin{figure*}[t!]
    \includegraphics{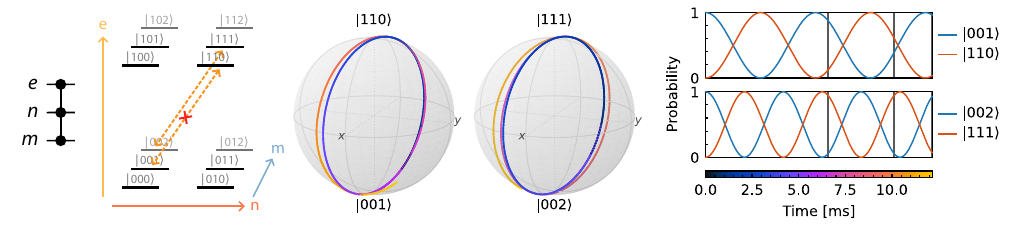}
    \caption{
        \textbf{Intra-quoct CCZ gate sequence}.
        The pulse sequence for the $e$-$n$-$m$ doubly controlled Z gate with fidelity $\mathcal{F}_{\text{CCZ}} \approx \fidenmccz$. As in Fig. \ref{fig:gate-seqs}, driven couplings between the relevant quoct states and the next-lowest motional states are shown, with those on which a $2\pi$ rotation are performed marked with a red X. Here, pulse sequences depend on the $n$ qubit because the driving light pulse is circularly-polarized.
    }
    \label{fig:gate-ccz}
\end{figure*}

Now we turn to operations that involve the motional states within the quoct space. A key observation to make when working in this space is that although motional states constitute robust quantum memories due to their insensitivity to their environments, transitions between motional states can only be driven indirectly; e.g. via some atomic transition through which a photon can change the atom's momentum. To achieve control over the population and phase of \textit{m}-qubit states, we work with the clock transition in the resolved-sideband regime where the Rabi frequency $\Omega$ is small compared to the frequency spacing between motional states $\omega$. This allows transitions between specific adjacent pairs of motional states to be individually addressed at the cost of extending gate times due to the reduced Rabi frequency. Here, we fix $\omega = 2\pi \times 100\,\text{kHz}$ as a reasonably large trap frequency attainable for scalable optical tweezer arrays.

The first gate operation we implement is a phase rotation to arbitrary angles on the $m$ qubit ($Z_m(\alpha)$). Operating in the frame co-rotating with the ground motional state, $Z_m$ can be implemented simply by adiabatically ramping the depth of the tweezer trap up and down such that the phase accrued by the $m = 1$ states is $\alpha$. We further construct pulse sequences wherein the red sideband on the clock transition is driven to implement an $e$-$m$ swap gate ($\text{SWAP}_{e,m}$), $e$-$m$ CZ gate ($\text{CZ}_{e,m}$), and $e$-$n$-$m$ doubly controlled Z gate ($\text{CCZ}$). Additionally, an $n$-$m$ swap gate ($\text{SWAP}_{n,m}$) can be implemented similarly to $\text{SWAP}_{e,m}$ by detuning one of each pair of beams used to drive stimulated Raman transitions within the $e = 0$ and $e = 1$ subspaces to a red sideband. Each pulse sequence comprises three separate pulses of identical Rabi frequencies and detunings resonant with said sideband but different durations and overall phases, such that the quoct state is rotated through an angle of $\pi$ on the appropriate transitions and axes, and $2 \pi$ on corresponding transitions to the set of next-lowest motional states (i.e. $m = 2$) as shown in Figs. \ref{fig:gate-seqs} and \ref{fig:gate-ccz}. Pulse sequences for each gate are found up to single $Z_e$, $Z_n$, and $Z_m$ gates by means of a gradient ascent-based optimization of the simulated gate fidelity. 

\begin{figure}[t!]
    \includegraphics[width=\linewidth]{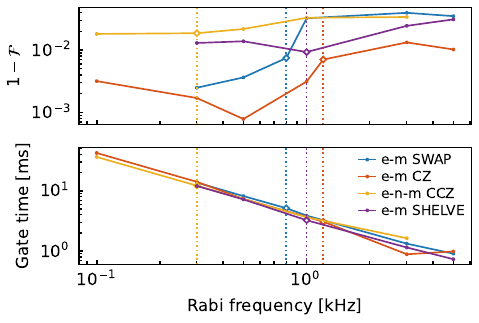}
    \caption{
        \textbf{Motion-selective gate fidelities and execution times}.
        Gate infidelities (top) and times (bottom) are shown as functions of Rabi frequency for each motion-selective gate implemented in this paper. Each point represents the maximum-fidelity condition for each gate that is discoverable via gradient ascent (see Appendix \ref{pulse-sequence}) for a particular Rabi frequency, with trap frequency fixed to $\omega = 2 \pi \times 100\,\text{kHz}$. Vertical dotted lines and diamond points show conditions selected for maximal fidelity at minimal gate time.
    }
    \label{fig:rabi-freq}
\end{figure}

Given our extensive use of clock transition sidebands, the Rabi frequencies $\Omega$ at which out pulse sequences are driven play a critical role in determining attainable gate fidelities. In Fig. \ref{fig:rabi-freq}, we plot expected optimum gate fidelities as functions of $\Omega$ and highlight chosen values of $\Omega$ for each gate [see Appendix \ref{pulse-sequence} for details]. At our chosen Rabi frequencies, we expect to reach gate fidelities of $\mathcal{F}_{\text{SWAP}_{e,m}} \approx \fidemswap$, $\mathcal{F}_{\text{CZ}_{e,m}} \approx \fidemcz$, $\mathcal{F}_{\text{CCZ}} \approx \fidenmccz$, and $\mathcal{F}_{\text{SHELVE}_{e,m}} \approx \fidemshelve$. In general, there is a correlation between gate fidelity and gate time. Hence, the fidelity of the CCZ could potentially be increased to the level of the others by slowing it down. Given this gate time/fidelity tradeoff, the optimal condition may depend on the algorithm being implemented. As seen below, the CCZ gate is needed only for the mass and chemical potential terms and is thus used relatively infrequently.

\subsection{Inter-quoct gates}
\label{inter}

Gates involving the states of two nearby atoms are commonly implemented through Rydberg blockade \cite{Isenhower2010, Saffman2010, Wilk2010, Jau2016, Graham2019, Levine2019}. In ququarts \cite{Jia2024}, single-photon transitions were driven from states in the $e = 1$ manifold to $6s\,ns\,\termsym{3}{S}{1}\,\ket{m_F = \pm 3/2}$ (effective principal quantum number $n^* \approx 55$) Rydberg states to implement a two-qubit CZ gate between the $e$ qubits of two nearby ququarts as well as a four-qubit CCCZ gate involving the $e$ and $n$ qubits of both ququarts. Crucially, the CZ and CCCZ gates were physically identical, up to only which Rydberg transitions were driven: While the CCCZ gate was realized by driving the $\ket{11} \leftrightarrow n^* \termsym{3}{S}{1} \ket{m_F = +3/2}$ transition in both ququarts, the CZ gate used a bichromatic pulse to simultaneously drive $\ket{10} \leftrightarrow n^* \termsym{3}{S}{1} \ket{m_F = -3/2}$ as well, thus making the CCCZ gate agnostic to the $n$ qubits of both ququarts.

\begin{figure}[t!]
    \centering
    \includegraphics[width=\linewidth]{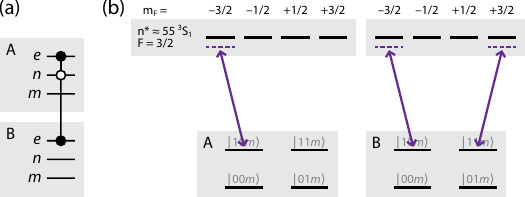}
    \caption{
        \textbf{Inter-quoct Rydberg-based C$\bar{\text{C}}$Z gate.}
        (a) The circuit symbol for a C$\bar{\text{C}}$Z gate on the $e$ and $n$ qubits of one atom and only the $e$ qubit of the other, with the $n$ qubit `anti-controlled'. (b) The driven Rydberg couplings to implement this gate. By driving both stretched transitions with a bichromatic pulse in quoct B, the gate can be made agnostic to quoct B's $n$ qubit.
    }
    \label{fig:inter-ccz}
\end{figure}

Here, we note that since we work with atoms at minimal temperatures--such that spatial wavefunctions will hardly change over the time scale of the Rydberg pulses--neither inter-atom gate identified above is sensitive to motion~\cite{Scholl2023b} and, hence, the inter-ququart CZ may be inherited without modification. However, our simulation requires the use of a three-qubit `C$\bar{\text{C}}$Z' gate where one qubit is `anti-controlled'; i.e. such that the $\pi$ phase shift is applied only when that qubit is in $\ket{0}$, rather than $\ket{1}$. This gate can be implemented natively as a gate involving the $e$ and $n$ qubits of one quoct and only the $e$ qubit of the other by applying the CCCZ gate pulse to the former with the drive on the $\ket{11m} \leftrightarrow n^* \termsym{3}{S}{1} \ket{m_F = +3/2}$ transition replaced with one on $\ket{10m} \leftrightarrow n^* \termsym{3}{S}{1} \ket{m_F = -3/2}$, and the bichromatic CZ gate pulse to the latter, as shown in Fig. \ref{fig:inter-ccz}. This results in a C$\bar{\text{C}}$Z gate that flips the phase of the $\ket{10m} \otimes \ket{1nm}$ two-quoct states and acts as the identity for all others. Note that this native implementation of the C$\bar{\text{C}}$Z gate requires the ability to address individual atoms with the relevant Rydberg pulses, however; if only global Rydberg pulses are available, this gate can alternatively be implemented by conjugating the two-quoct gate $I \otimes X_n$ with $(X_n \otimes I) \text{CCCZ}$.

\subsection{Readout protocol}
\label{readout}

\begin{figure*}[t!]
    \includegraphics{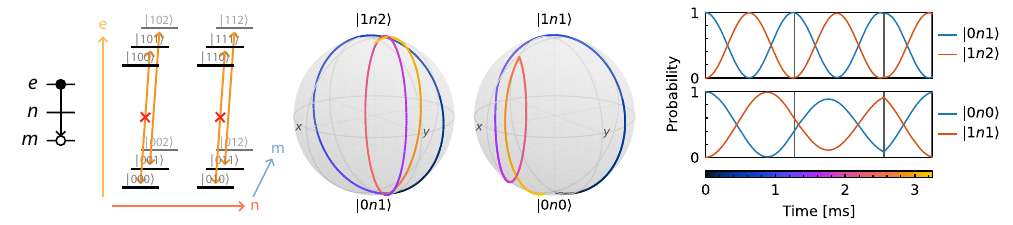}
    \caption{
        \textbf{Motional-selective shelving sequence for readout}.
        Pulse sequence implementing our $e$-$m$ shelving operation, where we take $\ket{0n1} \leftrightarrow \ket{1n2}$ while performing a simultaneous $2\pi$ rotation on $\ket{0n0} \leftrightarrow \ket{1n1}$ with fidelity $\mathcal{F}_{\text{SHELVE}_{e,m}} \approx \fidemshelve$. As in Fig. \ref{fig:gate-seqs}, we show simulated Bloch sphere trajectories and flattened probability time traces. We denote this operation by the circuit diagram symbol on the left for convenient use in Fig. \ref{fig:measurement}.
    }
    \label{fig:shelving}
\end{figure*}

Readout of the quoct space is performed using a multi-round procedure, wherein multiple measurements are performed within subsets of the full computation space. As for ququarts, photons are scattered indistinguishably from all $e = 0$ states with additional operations between each round, yielding a single classical bit per round corresponding to the $e$ qubit at the time of measurement: If photons are detected, the readout is said to have yielded a ``bright'' outcome ($e = 0$), and ``dark'' ($e = 1$) otherwise. Since we have implementations of both $\text{SWAP}_{e,n}$ and $\text{SWAP}_{e,m}$ available and each round of photon scatter effectively measures the $e$ qubit by mapping $e = 0$ ($1$) to a ``bright'' (``dark'') signal on a detector, full readout of the entire space should, in principle, be achievable in three rounds by merely swapping the $n$ and $m$ qubits into the $e$-qubit position between rounds. However, the photon scattering process couples incoherently to the motional degree of freedom, scrambling information stored in the $m$ qubit across a wide range of motional states but crucially leaving the $n$ qubit intact.

\begin{figure}[t!]
    \includegraphics[width=\linewidth]{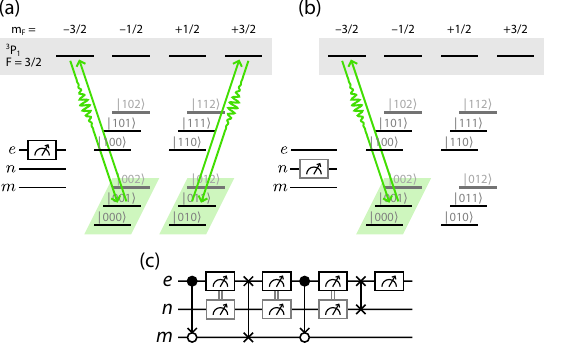}
    \caption{
        \textbf{Measurement protocol}.
        (a) A single round of $e$-qubit readout. The two stretched $\termsym{1}{S}{0} \leftrightarrow \termsym{3}{P}{1}$ are driven simultaneously, scattering photons that are indistinguishably detected to effectively measure exactly the $e$ qubit. The $n$ and $m$ qubits are only affected if $e=0$. (b) An alternative measurement process, where only the stretched transition to $\termsym{3}{P}{1} m_F = -3/2$ is driven, amounting to partial readout of the $n$ qubit, only within the $e = 0$ subspace. This measures $|00m\rangle$ or NOT $|00m\rangle$. (c) The circuit diagram for full readout of the quoct computational space. Each full readout requires a maximum of four rounds of $e$-qubit measurement, but can be short-circuited by a partial readout of the $n$ qubit as in (b) following a ``bright'' result from the $e$-qubit readout as indicated by the classically controlled, gray-boxed $n$-qubit measurements. See Appendix \ref{measurement-protocol} for more information.
    }
    \label{fig:measurement}
\end{figure}

Given some $N$ rounds of readout, we require a procedure capable of mapping classical bitstrings of length $N$ (the outcomes of each round of readout) surjectively onto the space of quoct basis states. Although motional scrambling prevents the use of naive swapping of the $n$ and $m$ qubits into and out of the $e$ qubit, we find that complete readout is possible through the use of an ``$e$-$m$ shelving'' operation constructed to swap a subset of basis states outside the computational space. Specifically, we use the blue sideband on the clock transition to transfer population from the $\ket{0n1}$ states to $\ket{1n2}$ as shown in Fig. \ref{fig:shelving}, and construct a pulse sequence to do so such that the accompanying $\ket{0n0} \leftrightarrow \ket{1n1}$ transition is driven through a net $2\pi$ rotation angle. As with the gates discussed in Section \ref{intra}, this operation is implemented using another three-part pulse sequence, the exact parameters of which were found using a gradient ascent optimization of the gate fidelity. In the context of readout and motional scrambling, the $e$-$m$ shelving operation can be seen as a way to move some information out of the subspace in which the scrambling occurs, such that no two states are scrambled together.

To then construct a full readout protocol, we employ a genetic algorithm to search the space of quoct circuits comprising a number of readout operations with any intra-quoct gates in between. Specifically, our algorithm minimizes the number of readout and intra-quoct gate operations needed to construct a protocol that performs the requisite surjective mapping from length-$N$ bitstrings to quoct basis states [see Appendix \ref{measurement-protocol} for details], and show one such protocol in Fig. \ref{fig:measurement}. This protocol is not unique; we find several others that are equivalent in terms of the number of rounds of readout and total gate count. However, we find none that satisfy the bitstring surjectivity constraint using fewer than four rounds of readout, and none that injectively map initial quoct states to final, output states after the protocol ends. These can be seen as artifacts due to, respectively, the shelving operation reducing the amount of information measured by each readout, and motional scrambling eventually mixing subsets of states together through multiple rounds of readout.

In Fig. \ref{fig:measurement}(b), we also identify an alternative measurement operation where photons are scattered from only the $\ket{00m}$ states. This operation was implemented experimentally in Refs. \cite{Huie2023,Norcia2023} and allows for direct measurement of the $n$ qubit within the $e = 0$ subspace. Here we note that, provided the ability to do real-time feedback during the execution of a circuit, it can be used to short-circuit full quoct readout in the case that an $e = 0$ result is detected from any of the $e$-qubit readouts: If photons are scattered during an $e$-qubit readout after shelving the $\ket{0n1}$ states, then the quoct collapses into a mixture of the $\ket{0n0}$ states. Hence, a single $n$-qubit readout using the alternative readout operation described above will then be enough to fully identify a single quoct basis state. In some cases, the readout circuit shown in Fig. \ref{fig:measurement}(c) can be reduced to only two rounds: If the first $e$-qubit readout yields $e = 0$, an $n$-qubit readout can be performed immediately after and the entire procedure subsequently terminated. See Appendix \ref{measurement-protocol} for a more thorough analysis of the possible readout results.

\section{Lattice QCD simulation in 1+1D}
\label{qcd}

\subsection{Lattice Hamiltonians to qubitized hardware}
\subsubsection{Kogut-Susskind Hamiltonian in the axial gauge}
\label{axial-gauge}We start with a review of the Kogut-Susskind (KS) Hamiltonian. The KS Hamiltonian for a three-color fermion $\psi = (\psi_r,\psi_g,\psi_b)^T$ in $1+1$ spacetime dimensions with one flavor takes the form \cite{Kogut1975,Sala2018,Farrell2023,Illa2024}
\begin{align}\begin{split} \hat H &= \frac{1}{2a} \sum_{n=0}^{2L-2} \left(\hat \psi_n^\dagger \hat U_n \hat \psi_{n+1} + \operatorname{h.c.}\right) + m\sum_{n=0}^{2L-1}(-1)^n \hat \psi_n^\dagger \hat \psi_n \\
&+ \mu\sum_{n=0}^{2L-1}\hat \psi_n^\dagger\hat \psi_n + \frac{ag^2}{2} \sum_{n=0}^{2L-2} \hat E_n^2
\end{split}
\end{align}
where $a$ is the lattice spacing, and $m,\mu,g$ are the bare mass, baryon chemical potential, and coupling strength respectively. The fermions are staggered so that the $L$ two-component spinor degrees of freedom are split into single-component fermions which live on the even and odd sites of a $2L$-site lattice (see Fig.~\ref{fig:overview}). On even sites excitations are understood as quarks $\hat \psi_n = (\hat c_r, \hat c_g, \hat c_b)^T$, and on odd sites as antiquarks $\hat \psi_n = (\hat c_{\bar r}, \hat c_{\bar g}, \hat c_{\bar b})^T$. $\hat U_n = \exp(-iag\hat A_n)$ is the Wilson line which transports $SU(3)$ color charge between site $n$ and $n+1$, and lives on the link. In the continuum limit $A_\mu\in \mathfrak{su}(3)$ is the gauge field which generates quantum chromodynamics akin to how the $A_\mu \in [0,2\pi) $ in QED generates the $U(1)$ gauge group. $\hat E^2_n$ is the $SU(3)$ Casimir which measures color-electric flux and also lives on the link.

For spatial dimensions $d > 1$, the KS Hamiltonian contains a color-magnetic term which introduces dynamics into the gauge field. In its absence, there are no independent gauge degrees of freedom. Fixing the axial gauge $A_x = 0$ and solving the Gauss law constraint for $A_t$, the Hamiltonian becomes
\begin{align}\begin{split} \hat H &= \frac{1}{2a} \sum_{n=0}^{2L-2} \left(\hat \psi_n^\dagger \hat \psi_{n+1} + \operatorname{h.c.}\right) + m\sum_{n=0}^{2L-1}(-1)^n \hat \psi_n^\dagger \hat \psi_n \\
&+ \mu\sum_{n=0}^{2L-1}\hat \psi_n^\dagger\hat \psi_n + \frac{ag^2}{2} \sum_{n=0}^{2L-1} \sum_{a=1}^8\left(\sum_{m=1}^n \hat Q_m^a\right)^2\label{axial-KS}
\end{split}
\end{align}
where $\hat Q_n^a = \hat \psi_n^\dagger T^a \hat \psi_n$ represents the color charge at site $n$ which preserves Gauss' law. Here we work with open boundary conditions (OBC) and so ignore topological effects. The continuum limit is discussed in Appendix \ref{axial-gauge-continuum}.

Observe that we have exchanged our local bosonic degrees of freedom $\hat E^2_n$ with nonlocal charge-pairing terms like $\hat Q^a_m\hat Q^a_n$ which decompose into fermionic degrees of freedom. Now that there are only fermionic degrees of freedom in our theory, we can introduce a finite-dimensional Hilbert space at each site. For the even sites we start with the unoccupied state $\ket 0$ and then raise by $\hat \psi_n^\dagger = (\hat c_r^\dagger, \hat c_g^\dagger, \hat c_b^\dagger)$, yielding $8=2^3$ possible states. As fermionic fields, the $\hat c_c$ satisfy the anticommutation relations $\{\hat c_c, \hat c_{c'}^\dagger\} = \delta_{c,c'}$ and $\{\hat c_c, \hat c_{c'}\} = \{\hat c_c^\dagger, \hat c_{c'}^\dagger\} = 0$. Therefore a fixed antisymmetric ordering of the states must be chosen. For example labeling the quarks on the \textit{same} site $q_1$ and $q_2$, we define for the diquark states
\[\ket{br} := \ket{q^b_1 q^r_2} - \ket{q^r_1 q^b_2} = -\left(\ket{q^r_1 q^b_2} - \ket{q^b_1 q^r_2}\right) = -\ket{rb}\]
Since there is a unique antisymmetric state, we do not need to create superpositions of states once we encode in the quantum computer; we can label the states as combinations of color excitations and then interpret them as the correct antisymmetric wavefunction. This gives us the quoct or ``qu8it" whose formalism was first developed in \cite{Illa2024} and we will review in part here:
\begin{widetext}
\begin{align}
\{\ket{\operatorname{quoct}} \}= 
\begin{array}{cccccccc}
\{ \ket0, & \hat c_r^\dagger \ket0, & \hat c_g^\dagger \ket0, & \hat c_b^\dagger \ket0, & \hat c_g^\dagger \hat c_b^\dagger \ket0, & \hat c_r^\dagger \hat c_b^\dagger \ket0, & \hat c_r^\dagger \hat c_g^\dagger \ket0, & \hat c_r^\dagger \hat c_g^\dagger 
\hat c_b^\dagger \ket0\}\\
\{ \ket0, & \ket r, & \ket g, & \ket b, & \ket {gb}, & \ket {rb}, & \ket {rg}, & \ket{rgb}\}\\
\end{array}
\end{align}
\end{widetext}
The ground state for the Dirac field is the Fermi sea with all the negative energy states occupied. The $\hat \psi_n = \hat c_{\bar c,n} $ operator on odd sites then produces antiparticles by lowering the occupied state. We follow \cite{Illa2024} and make the particle-hole transformation $\hat c_{\bar c} \rightarrow \hat c_c^\dagger$ so that quoct excited states correspond to the creation of antiquarks. The KS Hamiltonian can then be written in the basis of the quoct states. This amounts to a representation of operators $\hat O$ as matrices $\tilde O = [\hat O]_\text{quoct}$.

\textbf{Kinetic term:} The kinetic Hamiltonian is
\begin{align*}
\hat H_\text{kin} &= \frac12 \sum_{c}^{r,g,b}\sum_{n=0}^{2L-2}(-1)^n\left(\hat c^\dagger_{c,n} \hat c_{c,n+1}^\dagger + \operatorname{h.c.}\right).
\end{align*}
To construct the matrix representations of the ladder operators (denoted $\tilde c_c$ or $[\hat c_c]_\text{quoct}$) in the single-site Hilbert space of each quoct, we examine how each operator acts on the quoct basis:
\begin{align}\begin{split}
\tilde c_r &= \ket{0}\bra{r} + \ket{g}\bra{rg} - \ket{b}\bra{rb} + \ket{gb}\bra{rgb}\\
\tilde c_g &= \ket{0}\bra{g} -  \ket{r}\bra{rg} + \ket{b}\bra{gb} + \ket{rb}\bra{rgb}\\
\tilde c_b&= \ket{0}\bra{b} + \ket{r}\bra{rb} - \ket{g}\bra{gb} + \ket{rg}\bra{rgb}\label{annihilation-operators}
\end{split}\end{align}
Observe that the minus signs come from the choice of fermion ordering for the basis and satisfy anticommutation relations on the single site.\footnote{If we chose $\ket{br}$ instead of $\ket{rb}$, for example, there would be no minus phase in $\tilde c_r$ but two in $\tilde c_g$ and $\tilde c_b$.} As in the Jordan-Wigner transformation, we need to include a phase dependent on the number of occupied fermions in order to satisfy anticommutativity on different sites. On each site,  define $(-1)^{n_q}$ which measures the number of fermions present; in the quoct basis, we define the matrix $\tilde P = \operatorname{diag}(1,-1,-1,-1,1,1,1,-1)$ on each site.  Thus the operator $\hat c^\dagger_{c,n}$ gets mapped to the quoct string $\underbrace{\tilde P \otimes ... \otimes \tilde P }_{n-1} \otimes \tilde c_c^\dagger$ where the site label on the operator determines the size of the string. Thus
\[[\hat c^\dagger_{c,n} \hat c^\dagger_{c,n+1}]_\text{quoct} =  \tilde I ... \tilde I\otimes \left(\tilde c^\dagger_c \tilde P\right)\otimes \tilde c_c^\dagger \otimes \tilde I...\tilde I\]
Then the matrix form of the kinetic Hamiltonian is (suppressing single-site identity factors):
\begin{align}
\tilde H_\text{kin} &= \frac12 \sum_{c}^{r,g,b}\sum_{n=0}^{2L-2}(-1)^n\left[(\tilde c^\dagger_{c} \tilde P)_n \tilde c_{c,n+1}^\dagger - (\tilde c_{c}\tilde P)_n\tilde c_{c,n+1}\right]
\end{align}
\textbf{Mass and Chemical Potential:} The mass term is
\[\hat H_\text{mass} = m\sum_c^{r,g,b}\sum_{n=0}^{L-1} \hat c^\dagger_{c,n} \hat c_{c,n}\]
and in the quoct basis
\[[\hat c^\dagger_c \hat c_c]_\text{quoct}= \tilde M =  \operatorname{diag}(0,1,1,1,2,2,2,3)\]
The chemical potential will have an extra sign factor $(-1)^n$ that computes the difference in occupation numbers on even (fermionic) and odd (antifermionic) sites.
\[\hat H_\text{chem} = \mu\sum_c^{r,g,b}\sum_{n=0}^{L-1}(-1)^n \hat c^\dagger_{c,n} \hat c_{c,n}\]
\begin{align*}
    \tilde H_\text{mass} + \tilde H_\text{chem} = \sum_{n=0}^{L-1} (m + (-1)^n\mu)\underbrace{\tilde I ... \tilde I\otimes \tilde M \otimes \tilde I ... \tilde I}_{n}
\end{align*}
\textbf{Electric term: }We can write the charge operator on even sites in terms of the creation and annihilation operators as discussed in Appendix \ref{axial-gauge-continuum}:
\[\hat Q^a = \sum_{c = r,g,b}\sum_{c'=r,g,b} \hat c_c^\dagger T^a_{c,c'} \hat c_{c'}\]
From this it is clear that $\ket 0$ is annihilated and single quark states transform by $T^a$. What about the three diquark and one triquark states? From an algebraic perspective, $\ket 0$ is a color singlet and transforms under the trivial, one-dimensional representation ($\mathbf 1$) and the color-charge generator is $0$. Likewise the single quark states $\ket c$ transform under the three-dimensional fundamental representation ($\mathbf 3$) so the color-charge generator is $T^a$. The diquark states come from the tensor decomposition of $\mathbf 3\otimes \mathbf 3 = \mathbf 6 \oplus \mathbf {\bar 3}$. Since our quark states have to be antisymmetric, the diquark states must then transform under the three-dimensional antifundamental representation ($\mathbf{\bar 3}$) and their color-charge generator is $\bar T^a := -(T^a)^*$. Lastly the triquark states come from $\mathbf3 \otimes \mathbf 3 \otimes \mathbf 3 = \mathbf {10} \oplus \mathbf 8\oplus \mathbf 8 \oplus \mathbf 1$ and only the $\mathbf 1$ is totally antisymmetric so their color-charge generator is 0. This can also be computed directly by applying $\hat Q^a$ to a color state and using the fact that $T^a$ is hermitian and traceless. Thus from computing how $\hat Q^a$ acts on quoct states, we can write the matrix representation $\tilde Q^a$ in the quoct basis:
\[[\hat Q^a]_\text{quoct} = \tilde Q^a = \begin{pmatrix} 0 & 0 & 0 & 0 \\ 0 & T^a & 0 & 0 \\ 0 & 0 & \bar T^a & 0 \\ 0 & 0 & 0 & 0\end{pmatrix}.\]

On odd-sites, the charge operator has the following form:
\[\hat Q^a = \sum_{c = r,g,b}\sum_{c'=r,g,b} \hat c_{\bar c}^\dagger T^a_{\bar c,\bar c'} \hat c_{\bar c'}\]
So on odd sites we have a different matrix representation for the charge operator:
\[[\hat Q^a]_\text{quoct} = \tilde{\bar Q}^a = \begin{pmatrix} 0 & 0 & 0 & 0 \\ 0 & \bar T^a & 0 & 0 \\ 0 & 0 & T^a & 0 \\ 0 & 0 & 0 & 0\end{pmatrix}\]

Since the charge operators on different sites commute, the electric term of the Hamiltonian can be written
\begin{align*}\hat H_\text{elec} &= \frac{ag^2}{2}\sum_{n=0}^{2L-2}\left(\sum_m^n \hat Q_m^2 + 2\sum_{a=1}^8\sum_{m < m'}^n \hat Q^a_m\hat Q^a_{m'}\right)\\
&= \frac{ag^2}{2}\sum_{n=0}^{2L-2}(2L-1-n)\hat Q_n^2\\
&+ ag^2\sum_{a=1}^8\sum_{n=0}^{2L-2}\sum_{m > n}^{2L-2} (2L-1-m)\hat Q^a_n\hat Q^a_m
\end{align*}
On both even and odd sites, the operator $\hat Q^2_m = \sum_a \hat Q^a_m \hat Q^a_m$ has a simple site-local representation
\[\tilde Q^2 = (\tilde Q^a)^2 = (\tilde{\bar Q}^a)^2 = \frac 43 \operatorname{diag}(0,1,1,1,1,1,1,0)\] 
The charge pairing terms $\hat Q^a_m\hat Q^a_{m'}$ will be discussed more in section \ref{L>1} but a useful fact is that  $\tilde Q^a \otimes \tilde{\bar Q}^a = \tilde{\bar Q}^a \otimes \tilde Q^a$ and $\tilde Q^a \otimes \tilde Q^a = \tilde {\bar Q}^a \otimes \tilde {\bar Q}^a$. Thus we can rewrite the quoct representation of $\hat H_\text{elec}$ (suppressing identities)
\begin{align*}
    \tilde H_\text{elec} &= \frac{ag^2}{2}\sum_{n=0}^{2L-2}(2L-1-n)\tilde Q_n^2\\
&+ ag^2\sum_{a=1}^8\sum_{m=0}^{2L-2}\sum_{m' > m}^{2L-2} (2L-1-m')\tilde  Q^a_m \tilde Q^a_{m'}\\
&+ ag^2\sum_{a=1}^8\sum_{k=0}^{2L-2}\sum_{k' > k}^{2L-2} (2L-1-k')\tilde  Q^a_k \tilde {\bar Q}^a_{k'}
\end{align*}
where the second sum is over $m,m'$ both even or both odd, and the third sum is over $k,k'$ where either $k$ is even and $k'$ is odd, or vice versa.

\subsubsection{Translating quoct operators to hardware gateset}
\label{vacuum-persistence}
So far we have reviewed how to write operators $\hat O$ on the lattice as matrices $\tilde O = [\hat O]_\text{quoct}$ in the quoct Hilbert space. The main new work here is translating operators on the abstract quoct basis into circuit gates acting on the electronic, nuclear, and motional degrees of freedom $[\hat O]_{enm}$ and then exploring NISQ and near-future observables. 

For $L=1$ with one quark and one anti-quark site, the quoct matrix representation for the Hamiltonian is
\begin{align}\begin{split}\tilde H &= \frac{1}{2a}\sum_c\left((\tilde 
c^\dagger_c\tilde P)\otimes \tilde c^\dagger_c - (\tilde c_c \tilde P)\otimes \tilde c_c\right) + \frac{ag^2}{2} \tilde Q^2\otimes \tilde I\\
&+ m\left(\tilde M\otimes \tilde I + \tilde I \otimes \tilde M\right)
+ \mu\left(\tilde M \otimes \tilde I - \tilde I \otimes \tilde M\right)
\end{split}\end{align}
There is flexibility when encoding this representation as excitations in the $\ket{enm}$ basis. The simplest assignment maps the quark colors directly so that red, green, and blue excitations correspond directly to the electronic, nuclear, and motional degrees of freedom, respectively. Under this encoding, the basis states are reordered to follow the usual computational basis $\ket{000}, \ket{001},...,\ket{111}$. This simple assignment has the  benefit of clarifying how the circuit should be structured. The circuit is broken into four parts for each term in the Hamiltonian, and is Trotterized to form the full circuit.

\begin{figure}[t!]    \includegraphics[width=1\linewidth]{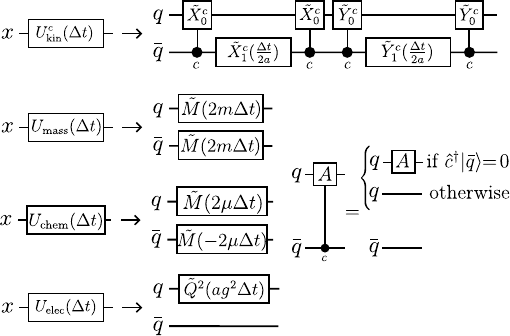}
    \caption{
        \textbf{Trotter decomposition for the $L = 1$ Hamiltonian}. For a given quoct operator $A$, we define $A(\alpha) := \exp(-\frac{i\alpha}{2} A)$ for the unitary operator describing its exponentiation. We also make use of a generalized controlled operator $CA$ which controls on a family of states.
    }
    \label{fig:trotter-decomp}
\end{figure}

\begin{figure*}[t!]
    \includegraphics[width=\linewidth]{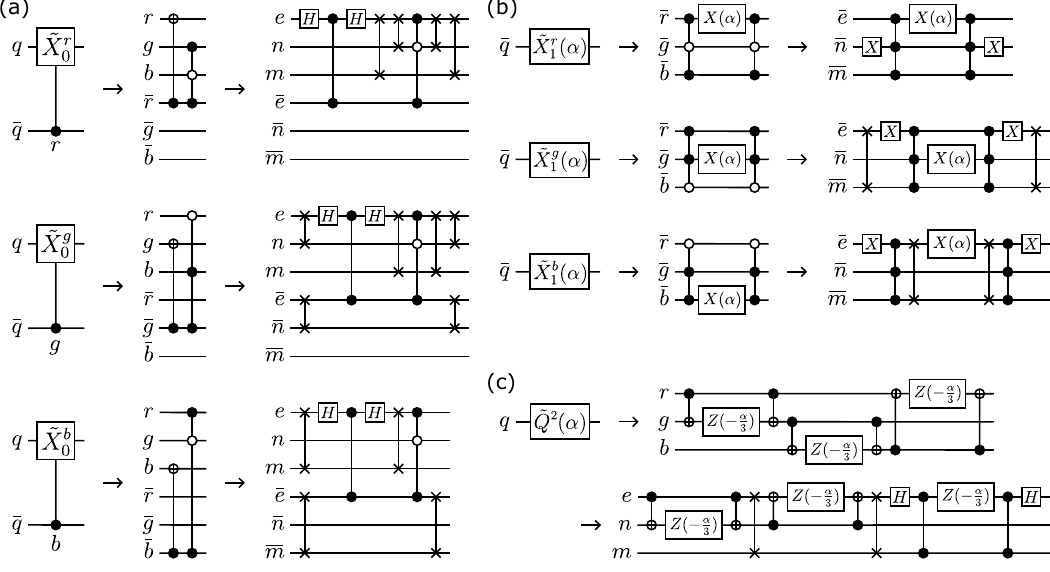}
    \caption{
        \textbf{Circuit compilation for the $L = 1$ Hamiltonian}. Each circuit is first understood as acting on quocts, then in terms of operators acting on each color sector, and lastly transpiled into the desired gateset.
    }
    \label{fig:circuits}
\end{figure*}

\textbf{Kinetic term:} For the kinetic Hamiltonian, we can decompose the creation operators into intra-quoct gates. For example looking at Equation \ref{annihilation-operators} we read off
\[\tilde c^\dagger_{r} = \frac12(X-iY)_{r}\otimes \operatorname{CZ}_{\bar gb}\] 
where $\operatorname{CZ}_{\bar gb}$ introduces a phase when $\ket b$ is occupied but $\ket g$ is not. Likewise $\tilde P = Z_r Z_g Z_b$. Thus for the red part of the kinetic term,
\begin{align*}
\big(\tilde c^\dagger_{r} \tilde P\big)_0 \tilde c^\dagger_{r,1} - \big(\tilde c_{r} \tilde P\big)_0 \tilde c_{r,1} &= \frac12 (X_0X_1 - Y_0Y_1)_r(\operatorname{CZ})_{g\bar b,0}\operatorname{CZ}_{\bar gb,1}\\
&=\frac12(\tilde X^r_0 \tilde X^r_1 - \tilde Y^r_0 \tilde Y^r_1)
\end{align*}
where
\[\tilde X^r_0 = X_r\operatorname{CZ}_{g\bar b} = \vcenter{\hbox{\includegraphics[scale = 1]{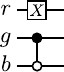}}}\]
$\tilde X ^r_1 = X_r \operatorname{CZ}_{\bar gb}$, and $\tilde Y^r_0$ and $\tilde Y^r_1$ have $Y$ gates replace $X$. We then find these circuit representations for the green and blue terms of the kinetic Hamiltonian in a similar manner. Since $\tilde X^c_i$ and $\tilde Y ^c_i$ are unitary, hermitian, and exchange states, we can circuitize the associated time-evolution unitary using the identity
\begin{align}
\vcenter{\hbox{\includegraphics[width=0.8\linewidth]{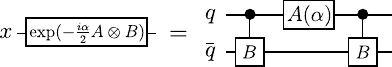}}}
    \label{circuit-identity}
\end{align}
where the controlled-$B$ quoct operator  represents a gate which is $B$ when $q$ is a certain family of states, and the identity $\tilde I$ otherwise, and $A(\alpha) = \exp(-\frac{i\alpha}{2}A)$ is the $SU(8)$ generalization of the usual rotations. This identity is proven in Appendix \ref{circuit-compilation}. Thus the kinetic term
\[\tilde H_\text{kin} = \frac1{4a}\sum_c \tilde X_0^c \tilde X_1^c - \tilde Y_0^c \tilde Y_1^c\]
exponentiates to the inter-quoct gate $U^c_\text{kin}(\Delta t)$ shown in Fig. \ref{fig:trotter-decomp}. 
Since we have chosen a basis for the quoct in terms of $(r,g,b)$ and $(e,n,m)$, we can write down the circuit for the controlled $C\tilde X_0^c$ and rotating $\tilde X_1^c(\alpha)$. This is demonstrated in Fig. \ref{fig:circuits}. The $C\tilde X^c_0$ is straightforward since it has a single red control (which controls half of the states). $\tilde X^c_1(\alpha)$ can be understood as a simple $X(\alpha)$ rotation with an extra phase produced by the $C\bar CZ$ gate. Observe that when transpiling to the optimal gateset for the hardware, native SWAP gates are often used to shuttle data between the motional and electronic / nuclear sectors. Thus while the the targets of the inter-quoct $C\bar CZ$ gate change in the color framework, they are all implemented by the same $C\bar CZ$ gate acting on the ququart $\ket{en}\otimes\ket{en}$ subspace of states shown in Fig. \ref{fig:inter-ccz}. More details for the intra- and inter-quoct gates are found in Section \ref{intra} and \ref{inter}.

\textbf{Mass and Chemical Potential:} The mass and chemical potential are always single intra-quoct gates as seen in Fig. \ref{fig:circuits}. Using the simple $(r,g,b)$ to $(e,n,m)$ mapping, we have that $\tilde M = 3/2 - \frac12 Z_r -\frac12 Z_g -\frac12 Z_b$. The constant term is irrelevant, and so $\tilde M$ exponentiates to the product of rotation gates $Z(\alpha)$.

\textbf{Electric term:} As noted for $L=1$, the electric term is a single intra-quoct gate. The charge operator has the simple form \[\frac{ag^2}{2}\tilde Q^2 = \frac{ag^2}{6}(3 - Z_rZ_g - Z_gZ_b - Z_rZ_b)\]
which transpiles to the $CX$ and $RZ$ circuit in Fig \ref{fig:circuits} in the usual way for Pauli strings.

The full circuit with some simplifications can then be found in Appendix \ref{circuit-compilation}. Note that while the direct $(r,g,b) \to (e,n,m)$ mapping makes developing the circuit simple, it is not necessary.

\subsection{Physical Observables}

\begin{figure}[t!]
    \centering
    \includegraphics[width=0.9\linewidth]{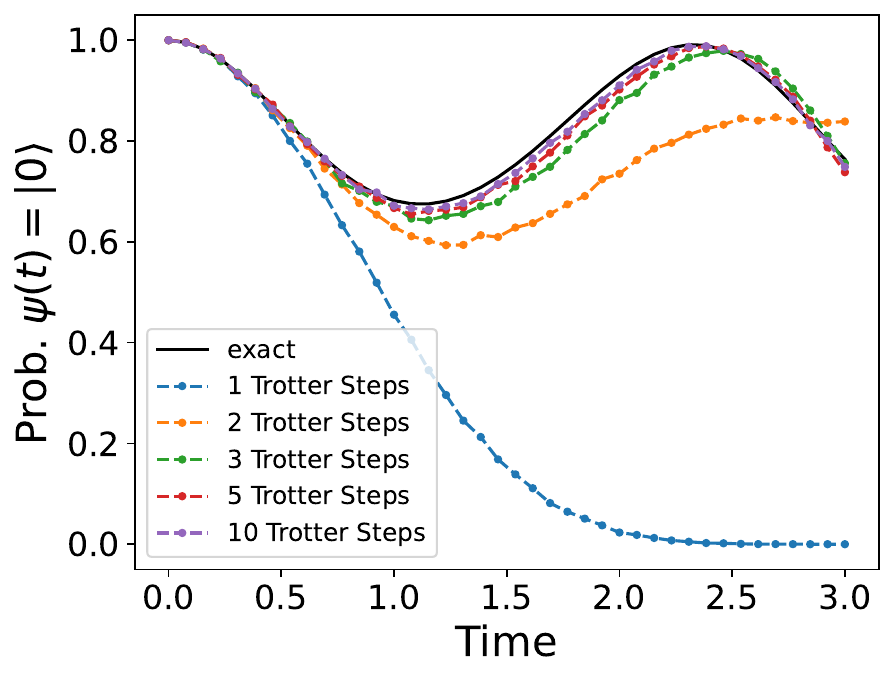}
    \caption{\textbf{Vacuum persistence for $L=1$}. Here we plot $|\braket{0|\psi(t)}|^2$ or the probability that $\psi(t) = \ket 0$. Each curve has a fixed number of Trotter steps so Trotter error grows for larger $t$. The fully unoccupied state oscillates into color-neutral multiparticle states for finite $g$. In this plot $a = m = g = 1$, $\mu = 0$. Convergence to the exact diagonalization (ED) result is seen for higher Trotter step density.}
    \label{fig:vac-persistence}
\end{figure}

\subsubsection{Vacuum persistence} 
A simple observable is vacuum persistence $|\braket{\psi(0)|\psi(t)}|^2 = |\braket{0|e^{-iHt}|0}|^2$, where the system is initialized in the unoccupied state and allowed to time evolve \cite{Farrell2023}. For finite $g$ the unoccupied state is not an eigenstate, and so the system will evolve nontrivially as seen in Fig. \ref{fig:vac-persistence}. Trotter error occurs as the different terms are not mutually commuting. Using first-order Trotterization the error is proportional to $t^2/D$ where $D$ is the number of Trotter steps. While one Trotter step is clearly insufficient as the pair creation terms of the kinetic term take the state away from the vacuum, we can see that three Trotter steps is sufficient to capture the oscillatory behaviour of the system at least until three time units.

\subsubsection{Adiabatic string breaking} 
\label{string-breaking}

\begin{figure}[t!]
    \includegraphics[width=0.9\linewidth]{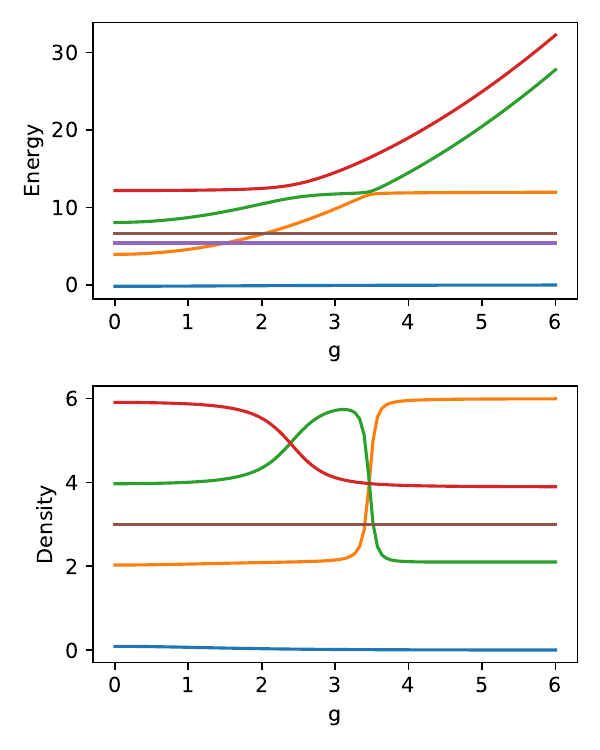}
    \caption{\textbf{Spectrum and expectation value of the fermion number density  from exact diagonalization for $L=1$}. Here we restrict to color-singlet states and fix $a=1, m=2, \mu = 0.2$, varying the coupling $g$. At $g=0$, the first excited state is a meson realized as a quark on the $n=0$ site, an antiquark at $n=1$, and a color-electric flux tube between them. At large $g$ this state transitions to a baryon $rgb$ and anti-baryon $\bar r \bar g \bar b$ with no flux between. At $g = 0$ and large $g$, the six states of the spectrum correspond to the ground state, the single baryon and anti-baryon state (split by the chemical potential $\mu$), and linear combinations of the meson $q\bar q$, two-meson $qq\bar q\bar q$, and the two-baryon state. Note that the transition of the single meson state (in orange) is accompanied by a small gap between it and the next excited state.
    }
    \label{fig:spectrum}
\end{figure}

In the $g\to\infty$ limit and ignoring color non-singlet states that may appear from the finite size of the lattice, we expect only the baryon states to have finite energy since in the lattice they have no color-electric flux. However at low $g$, the single meson state has lower energy since it has less mass. From exact diagonalization in Fig. \ref{fig:spectrum} we can see in the spectrum that this is a transition from a quark-antiquark state connected by a color fluxtube to a bound baryon-state, as noted in \cite{Farrell2023}. We can follow the transition of this excited state by adiabatically increasing or decreasing $g$.

For adiabatic evolution, we have an initial time-independent Hamiltonian $H_i$ with an eigenstate that we know well, $\ket{\psi_i}$, and we have a target Hamiltonian $H_f$ with less well-known eigenstates $\ket{\psi_f}$. Consider the time-dependent Hamiltonian 
\[H(t) = \left(1-\frac tT\right)H_i + \frac tT H_f\]
Then $H(0) = H_i$ and $H(T) = H_f$. We can transport our initial state by time evolution as $\ket{\psi_i(t)} = U(t) \ket{\psi_i(0)}$ where
\begin{align}\begin{split}U(T) = \mathcal Te^{\displaystyle -i\int_0^T H(t)dt} \approx \mathcal T \prod^{D}_{j=0} e^{\displaystyle-iH(j\Delta t)\Delta t}\\
\approx \underbrace{e^{\displaystyle -i H(T-\Delta t)\Delta t}...e^{\displaystyle -i H(\Delta t)\Delta t}e^{\displaystyle -i H(0)\Delta t}}_D
\end{split}\end{align}
and where $\Delta t = T/D \ll 1$ and $D$ records the depth of the circuit. 
Assuming that $T$ is long enough, then the adiabatic theorem tells us $\ket{\psi_i(T)} \cong \ket{\psi_f}$ up to a phase \cite{Tameem2018}. 

We can rewrite $H(t)$ for our particular Hamiltonian,
\[H_g = H_K + H_M + g^2 H_E\] 
Then
\[U(T) = \mathcal T\prod^{D}_{j= 0} e^{\displaystyle-iH_{g_\text{eff}}\left(\frac j D\right)\Delta t}\]
where $g_\text{eff}^2 = g_i^2 + (j/D)(g_f^2-g_i^2)$
and where $\Delta t$ is small enough to approximate the time-ordered integration and allow for Trotterization.

\begin{figure}[t!]
    \centering
    \includegraphics[width=0.9\linewidth]{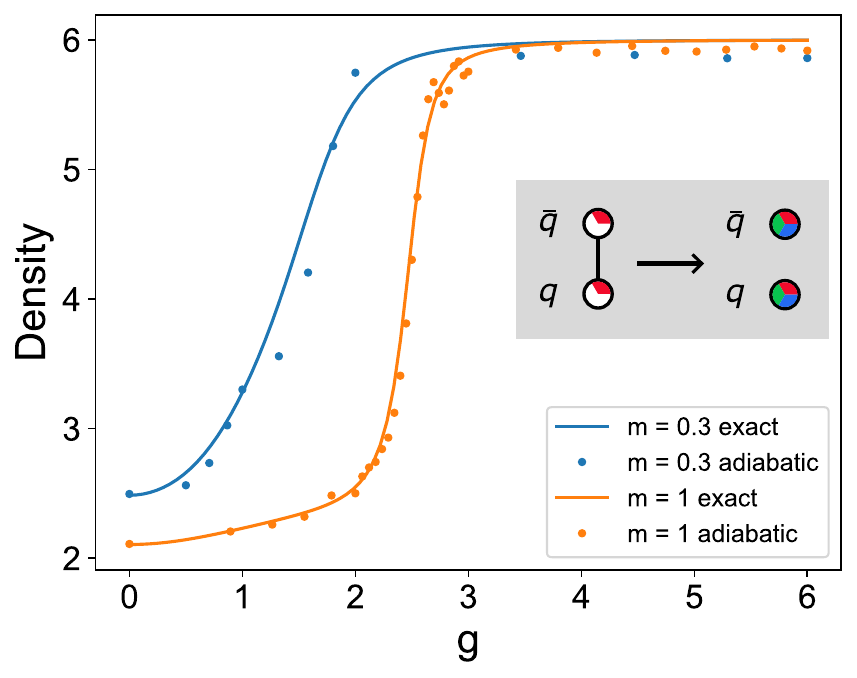}
    \caption{\textbf{Adiabatic evolution of the single meson state at weak coupling to the two-baryon state at strong coupling.} Here $L=a=1, \mu =0$. The evolution is broken into three Trotter ``speeds" depending on the spectral gap. In particular the transition in density from the single meson to the two baryon states is accompanied by a small gap between it and the next excited state, requiring more adiabatic time and thus more Trotter steps. Circuits are measured at unit intervals in $T$. For $am=1$ in orange, the maximum circuit depth is 250 Trotter steps. For $am = 0.3$ in blue the maximum circuit depth is 60 Trotter steps.}
    \label{fig:adiabatic2}
\end{figure}

For $g=0$ the eigenstates of the Hamiltonian are exactly solvable \cite{Farrell2023} for arbitrary $m,a,\mu$, and so the system can be initialized at this point and then in principle evolved to any intermediate strength $g$ even for systems that are not tractable on classical computers. An example of this is shown in Fig. \ref{fig:adiabatic2}. For larger $m$ the gap between excited states shrinks at the transition, and the $g=0$ and $g\to\infty$ states partition better into integer densities. For $L=1$, the single baryon states are split by nonzero $\mu$, but cannot evolve as $g$ changes, making their dynamics fairly trivial. For $\mu > m$, the lowest energy state is the single antibaryon as expected.

\subsubsection{Baryon size} 
\label{L>1}
\begin{figure}[t!]
    \centering
    \includegraphics[width=0.9\linewidth]{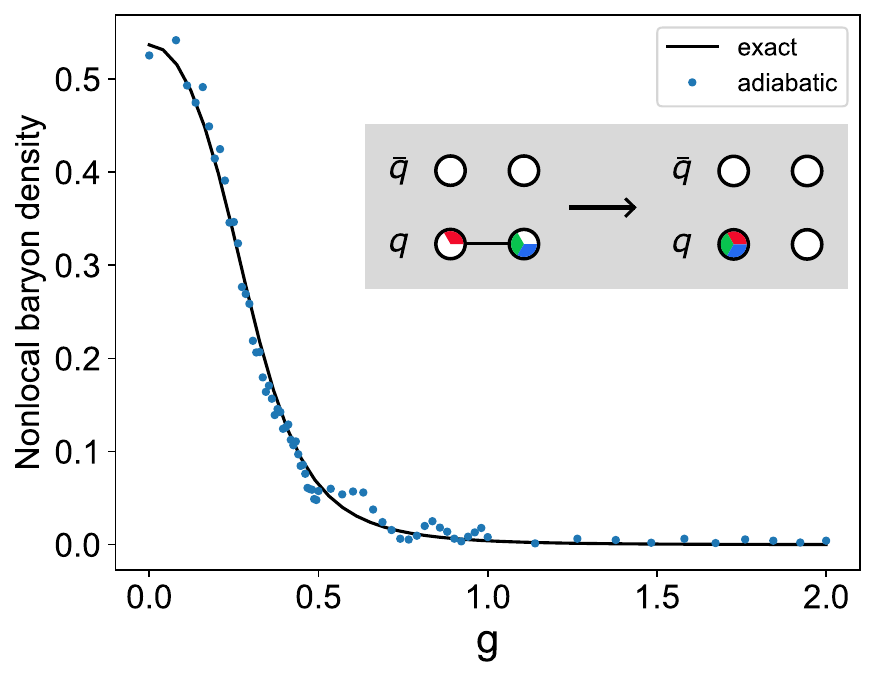}
    \caption{\textbf{Adiabatic evolution of the density of nonlocal baryons for $L=2,m=2,\mu = 0.1$}. For $L=2$ the baryon can be extended on to two lattice sites, e.g. $\ket r$ on site $0$ and $\ket {gb}$ on site $2$. As $g$ approaches unity, the baryon is localized to only one site. The maximum circuit depth is 500 Trotter steps.}
    \label{fig:delocal}
\end{figure}

As $L$ increases, the spectrum of physical states becomes richer. Baryon number $n_B$ is a symmetry of the Hamiltonian, and for $L=1$ the single baryon and anti-baryon states are fixed as $g$ evolves since there is only one state for each sector. For $L=2$ and finite $g$, there are $20$ states in the $n_B = 1$ sector which allows for the low-energy state pictured in Fig \ref{fig:delocal} to evolve as $g$ is increased. Along with single-baryon states on one site, the $n_B = 1$ eigenstates also have support on states with two quarks on one site and one quark on the other site, with a color-electric flux tube connecting them. Thus we have baryons which are ``nonlocal". The density of these states can be measured using the following operator 
\[(2 \text{ particles})_0\otimes (1 \text{ particle})_2 +  (1 \text{ particle})_0\otimes (2 \text{ particles})_2\]
which projects out these nonlocal states. 

Similar to the string-breaking observable, increasing $g$ means the flux-tube becomes energetically expensive. However for this low-energy state, rather than breaking the string by producing color-neutralizing partners, the baryon eliminates the flux-tube by huddling all the quarks on one site. Thus the total number density remains constant, but the local number density changes. These nonlocal baryons can be thought of as having a finite size relative to the lattice.

\subsection{Resource Estimation and $L>1$}
For $L=1$ the basic circuit for one Trotter step is shown in Fig. \ref{fig:full-circuit}. Based on the MPP ququart operations with fidelities $\gtrsim 0.999$ and the intraquoct gates with fidelties shown in Table \ref{tab:pulse-angles}, the overall fidelity of one Trotter step is $0.691$. Averaging the gate times to $\sim 1\,\text{ms}$, then each Trotter step takes $\approx 100\,\text{ms}$. Since coherence times are $\sim 1\,\text{sec}$, we can run $\lesssim 10$ Trotter steps within coherence times at a total fidelity of $0.025$. However as noted, the three Trotter step vacuum persistence circuit matches the ED results closely and would have a larger fidelity of $0.331$. Thus while the fidelity makes adiabatic evolution for observables like string-breaking and baryon locality challenging, it is clear that simple observables like vacuum persistence are achievable in the short-term.

It should be noted that the low number of inter-quoct gates is particular to $L=1$, as the electric term of the Hamiltonian is confined to one quoct as a $\tilde Q^2$ term. For $L > 1$ the electric term is nonlocal and generates $(2L-1$ choose $2)$ pairs of the type $\tilde Q\otimes \tilde{\bar Q}$ and $\tilde Q\otimes \tilde Q$. Thus while the other terms of the Hamiltonian scale linearly with $L$, the electric term grows quadratically. 

Moreover, the nonlocal charge terms do not have simple qubit or quoct circuit representations like the other terms. One way to represent them is to decompose the Hamiltonian term via Pauli string decomposition, and then diagonalize mutually commuting families to cut down on circuit complexity as done in \cite{Farrell2023}. It is simple to then write circuits which exponentiate these Pauli strings composed of $Z$ and $I$.

We can also examine these terms and try to write quoct operators directly. For example $\hat Q^1$ exchanges red and green quarks so it is built around a $\operatorname{SWAP}_{e,n}$ gate. Unlike $\tilde X_i^c$ and $\tilde Y_i^c$, however, 
\[\tilde Q^1 = \ket{100}\bra{010} + \ket{010}\bra{100} - \ket{101}\bra{011} - \ket{011}\bra{101}\]
is not unitary which makes its circuit representation once exponentiated less trivial. We can devise operations in terms of quocts and ququarts
\[\exp(i\alpha\tilde Q^1) = \vcenter{\hbox{\includegraphics[scale = 0.9]{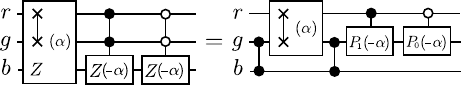}}}\]
where the boxed 3-qubit unitary represents a ``rotated" $\operatorname{SWAP}_{e,n} Z_m $ gate:
\[(\operatorname{SWAP}_{e,n} Z_m )(\alpha) = \cos\left(\frac{\alpha}{2}\right) I_{64} - i\sin\left(\frac{\alpha}{2}\right) \operatorname{SWAP}_{e,n} Z_m, \]
where the boxed 2-qubit unitary represents just the rotated $\operatorname{SWAP}_{e,n}$,   $P_1$ is the phase shift gate, and $P_0 = XP_1X$. If possible, these operations leverage the quoct architecture more efficiently.

Unfortunately, both methods generate color non-singlet states due to the non-commutativity of the charge operators ${[\hat Q^a_m, \hat Q^b_n] = i\delta_{mn}f^{abc}\hat Q^c_m}$ where $f^{abc}$ is the $SU(3)$ structure constant. Thus Trotter error will mix physical and non-physical states. One solution is to artificially increase the energy of non-singlet states by introducing a term proportional to the sum of the charge over the lattice squared $(\sum_{n=0}^{2L-1} \hat Q^a_n)^2$ as done in $\cite{Farrell2023,Illa2024}$. Then Gauss' law is reestablished by exponentially suppressing the nonphysical states. However this introduces $(2L$ choose $2)$ more of these charge pairing terms. A quoct operator that exactly exponentiates the charge coupling terms would resolve this.

Since new features of the theory are present in the simulation as $L$ increases, such as the finite size of the baryon in Fig. \ref{fig:delocal}, it is necessary to work on developing circuits which scale reasonably for larger $L$. Furthermore, larger scale simulations will be important for scattering simulations that are out of reach for classical simulation \cite{Jordan2012, Davoudi2024}.

\section{Outlook}
We have demonstrated a hardware-efficient approach to digital quantum simulation of nuclear physics with atomic qudits in which the higher-dimensional space within each atom is ``qubitized." We have demonstrated a universal gate set and have shown that two atoms (a quark site and an antiquark site) are sufficient to simulate oscillations between vacuum and quark-antiquark pairs and to simulate string breaking as the coupling strength is varied. We believe that our multi-qubit encoding scheme will contribute to a broad range of applications within quantum computing and digital quantum simulation. Finally, we believe that our use of motional states and our family of composite pulses to isolate the lowest two motional states will aid the development of hybrid digital/continuous variable and bosonic encodings for neutral atoms~\cite{Grochowski2023, Crane2024, Liu2024a, Liu2024b, Bohnmann2025}.

Although the resource requirements for the near-term realizations identified above are daunting, we highlight opportunities to mitigate or even correct errors. These opportunities are enabled by a ``physics-inspired" co-design for error detection/correction protocols that respect the underlying symmetries of the Hamiltonian. Specifically, for $L=0$ with the constraint of zero total baryon number, our system exhibits a matter-antimatter symmetry. Hence, the occupancy of a quark of any color guarantees the occupancy of an antiquark of the same color. Accordingly, a minimal strategy would be to post-select on the 8/64 physically meaningful states: $|rbg\bar{r}\bar{b}\bar{g}\rangle=\{|000000\rangle,|100100\rangle,|010010\rangle,|001001\rangle,|110110\rangle,|101101\rangle,\\
|011011\rangle,|111111\rangle\}$. While still useful, this approach drastically reduces the success rate and does not identify self-correcting errors that occur during the circuit. Thus, we propose to add color-by-color parity checks after each Trotter step. This could be accomplished with an ancilla qubit for each color, or potentially with an ancilla quoct that contains three ancilla qubits. This approach would keep a record of matter-antimatter parity errors during the circuit, which may enable classical decoding protocols to salvage those operations. Additionally, it may be possible to correct such errors with a feedforward operation that flips one of the two bits. However, since the energy of a $|00\rangle$ (for a $|c\bar{c}\rangle$ state) may be significantly different than that of $|11\rangle$, depending on $g$, not knowing which of the two to flip could lead to a substantial injection or depletion of energy.  

Nevertheless, we anticipate that ``physics-inspired" error detection/correction protocols that are co-designed with our multi-qubit-encoded hardware will open the door to robust and resource efficient simulation of nuclear physics phenomena, and we leave this study for future work. Given the advantages of using qudits to encode the many degrees of freedom of nuclear matter, we anticipate that optimal error detection/correction protocols should be based on qudit schemes rather than the canonical picture of error correcting qubit codes. Examples might include bosonic codes such as the Gottesmann-Kitaev-Preskill (GKP) code~\cite{Gottesman2001} or closely-related angular momentum-based codes~\cite{Albert2020,Jain2024}. It will be interesting to interpret the code words of such codes as robust states of quark matter. Our ``qubitization" of qudits and qubit-to-color mapping may aid in the co-design of error correcting codes inspired by the underlying nuclear physics. Alternatively, it could be interesting to fully embrace the quoct space without qubitizing, potentially offering a more direct translation to bosonic and qudit error correcting codes. 

Indeed, our ``qubitization" and qubit-to-color mapping is one of many possible methods to implement the KS Hamiltonian. Our large gateset suggests that it could be possible to directly operate in the quoct space, which could more readily enable extensions of our model, such as the use of Weyl gauge instead of axial gauge. For the specific model considered here (1+1D QCD in axial gauge with $L=1$), we do not believe that any other encoding scheme is more resource efficient, but we anticipate that the resource cost differential between different encoding schemes will begin to deviate as $L$ increases or as we go to higher dimensions. Nevertheless, our approach offers an intuitive and compelling route to efficiently study small-scale nuclear physics on near-term quantum hardware.\\

\textit{Acknowledgments}.---We acknowledge Tian Xue for stimulating discussions. J.P.C. acknowledges funding from the DOE Early Career Award (award number DE-SC0025655) from the Office of Nuclear Physics Quantum Horizons Program, and from the NSF (award number 2339487). P.D. acknowledges funding from the U.S. Department of Energy, Office of Science, Office of High Energy Physics Quantum Information Science Enabled Discovery (QuantISED) program.

\setcounter{section}{0}



\appendix
\renewcommand\appendixname{APPENDIX}
\renewcommand\thesection{\Alph{section}}
\renewcommand\thesubsection{\arabic{subsection}}



\setcounter{figure}{0}
\renewcommand{\thefigure}{S\arabic{figure}}

\section{Analysis of motion-preserving pulses}
\label{mpp-analysis}

Here, we analyze the effects of motion-preserving pulses (MPPs) \cite{Lis2023} in the context of the larger quoct computational space. MPPs are a specific use of the Compensation for Off-Resonance with a Pulse SEquence (CORPSE) scheme \cite{Cummins2003} where two CORPSEs, each implementing a net $90^\circ$ rotation on the relevant transition, are concatenated to perform a total rotation of $180^\circ$. The authors of Ref. \cite{Lis2023} found that this six-part composite pulse sequence also preserves the motional ground state and used it as a way to mitigate unwanted atomic heating when driving the clock transition: Even when the Rabi frequency $\Omega$ at which the clock transition is driven is much larger than the trap frequency $\omega$, the MPP evolves the atomic state through such a trajectory that an atom beginning in its motional ground state will be left with negligible overlap with higher motional states.

\begin{figure}[t!]
    \centering
    \includegraphics[width=\linewidth]{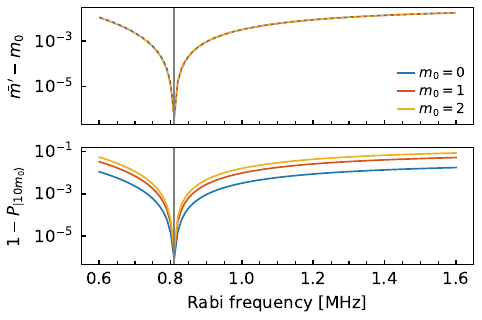}
    \caption{
        \textbf{Preservation of motional quantum numbers with MPPs.}
        Starting in the $\ket{00m_0}$ quoct state, we simulate a $X_e(\pi)$ using a MPP and plot the change in average motional quantum (top) and the $\pi$-pulse infidelity (bottom) as a function of the Rabi frequency $\Omega$ used for the gate for a trap frequency of $\omega = 2\pi \times 100\,\text{kHz}$. The change in motional quantum is nearly identical for all $m_0$, and we identify $\Omega_0 \approx 2\pi \times 810\,\text{kHz}$ as the optimum Rabi frequency for both quantities. We note that $\Omega_0$ agrees proportionally to that in Ref. \cite{Lis2023}, which found an optimum at $\approx 2\pi \times 80\,\text{kHz}$ for $\omega = 2\pi \times 10\,\text{kHz}$.
    }
    \label{fig:mpp}
\end{figure}

We find in simulation that the MPP sequence additionally preserves the $m = 1$ and $m = 2$ states, which allows it to be used in the full quoct space for more general-purpose operations. We also find that this behavior is maintained even at larger trap frequencies than those originally studied in Ref. \cite{Lis2023}. Whereas Lis \textit{et al.} work at a trap frequency of $2\pi \times 10\,\text{kHz}$, we fix to $\omega = 2\pi \times 100\,\text{kHz}$ so that larger Rabi frequencies may be used for the motion-selective operations discussed in Appendix \ref{pulse-sequence}. In Fig. \ref{fig:mpp}, we plot the change in average motional quantum and associated $\pi$-pulse fidelity for MPPs applied to states starting in an initial motional state of $m_0 \in \{0, 1, 2\}$ as a function of $\Omega$, and identify an optimum Rabi frequency $\Omega_0 \approx 2\pi \times 810\,\text{kHz}$. $\Omega_0$ is approximately equal in proportion to that in Ref. \cite{Lis2023}, where the authors found an optimum at $\approx 2\pi \times 80\,\text{kHz}$ for a trap frequency of $2\pi \times 10\,\text{kHz}$. Fixing the MPP Rabi frequency to $\Omega_0$, we find that MPPs preserve superpositions of motional states, up to a negligible phase shift ($\lesssim 1^\circ$) within the superposition, and a global shift of $\approx -88^\circ$ applied to all states involved the transition (i.e. that undergo dynamics due to the pulse). We note additionally that population transfer between the $e = 0$ and $e = 1$ subspaces can alternatively be achieved using a three-photon process (rather than driving the clock transition directly) \cite{Barker2016, Panelli2025} to reduce the requisite optical power for a $\approx 810\,\text{kHz}$ Rabi frequency.

\begin{figure}[t!]
    \centering
    \includegraphics[width=\linewidth]{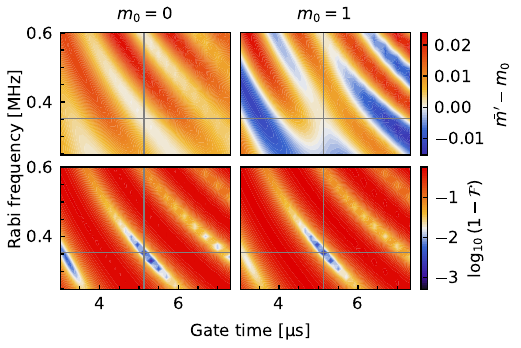}
    \caption{
        \textbf{Preservation of motional quantum numbers with Hadamard pulses.}
        The change in average motional quantum (top) and overall pulse infidelity (bottom) for initial motional state $m_0 = 0$ (left) and $m_0 = 1$ (right) as functions of Rabi frequency and gate time. The vertical and horizontal lines show values for both parameters that minimize the pulse fidelities, coinciding at the same values for both initial motional quanta.
    }
    \label{fig:hadamard}
\end{figure}

We similarly analyze the ququart-space pulse implementing the Hadamard gate from Ref. \cite{Jia2024} in the context of varied initial motional states and superpositions thereof. Unlike for MPPs, the Hadamard pulse protocol does not strictly require a certain total nutation angle: For a given Rabi frequency, the detuning of the pulse is chosen such that population inversion occurs between the single computational basis states coupled by the transition and the appropriate X-basis states thereof, while the gate time can be freely chosen such that the final state at the end of the pulse has greatest overlap with the expected X-basis state. As such, we again simulate the change in average motional quantum and overall pulse fidelity for initial motional quantum $m_0 \in \{0, 1\}$ ($m_0 = 2$ is excluded because it is only relevant to measurement), but now freely vary the gate time in addition to the Rabi frequency. The results are shown in Fig. \ref{fig:hadamard}. We find optimal Rabi frequency $\Omega_0 \approx 2\pi \times 350\,\text{kHz}$ and gate time $T_0 \approx 5.1\,\text{$\mu$s}$ for both $m_0 = 0$ and $m_0 = 1$. For this condition, we find that both the change in average motional quantum and the pulse infidelity are $\lesssim 10^{-3}$, with a phase shift of $\approx 175^\circ$ applied to all $m = 1$ states.

\section{Pulse sequence optimization}
\label{pulse-sequence}

The pulse sequences implementing intra-quoct two- and three-qubit gates presented in the main text [see Figs. \ref{fig:gate-seqs} and \ref{fig:shelving}] were found by means of a gradient ascent optimizer applied to simulated multilevel dynamics under applied laser pulses. While some approaches to the general problem of pulse optimization may optimize a time-dependent phase to implement a particular gate \cite{Jandura2022}, we choose to optimize a technically simpler sequence of only three concatenated pulses. Each pulse is on resonance with the relevant sideband of the clock transition, and our optimization scheme varies only the nutation and phase angles for each pulse. This can be seen as a simplified instance of the approach taken in Ref. \cite{Liu2021}, which expands the computational space to include higher motional modes as a general qudit.

Our scheme is based on the observation that the effective Rabi frequency at which a particular sideband transition will be driven (by a pulse of fixed power and detuning) varies with the motional quantum numbers of the states involved. This allows the motional degree of freedom to be controlled such that undesired leakage out of the computational space is prevented via sequences that perform a net-$2 \pi$ rotation on erroneous transitions. More generally, there is a trade-off in the number of pulses used in a sequence between the fidelity and duration of the overall operation: A larger number of pulses results in greater control over state trajectories, but longer gate times as well. Since we work in the resolved-sideband regime, the gate time associated with additional pulses in a sequence can be costly. In this regime, Rabi frequencies $\Omega$ are constrained to be much less than the trap frequency $\omega$. This introduces another factor in the trade-off between fidelity and gate time, where a larger Rabi frequency will reduce not only the gate time but motional selectivity as well. It follows that increasing $\omega$ will then also allow $\Omega$ to be increased, but we find an upper limit to this effect in practice. We plot the fidelity and gate time as functions of $\Omega$ for a realistic limiting trap frequency $\omega = 2 \pi \times 100\,\t{kHz}$ in Fig. \ref{fig:rabi-freq}. Here, we fix to sequences of three pulses and suitable Rabi frequencies to maintain gate fidelities of $\gtrsim 0.99$, with $\sim \t{ms}$ gate times.

We simulate the quoct space with motional quantum numbers up to $m = 2$, assuming equalization of nuclear splittings in the ground and clock manifolds via a suitable set of background light shifts \cite{Jia2024}. When simulating gates that only involve a subset of the encoded qubits, however, we reduce the size of the simulated Hilbert space to only the relevant degrees of freedom. Given a target unitary matrix $U_0$ describing a gate operation to implement, we construct an objective function $\mathcal{F}$ from the gate fidelity comparing $U_0$ to an effective unitary $U'$ computed for pulse angles $\{\theta_k\}, \{\varphi_k\}$ ($k \in \{0, 1, 2\}$) by initializing in each available basis state, simulating the pulse sequence, and taking the final states. The full objective function is then
\begin{equation}
    \mathcal{F} =
        \max_{\alpha, \beta, \gamma}
            \frac{1}{N} \Big|
                \t{Tr}\big(
                    U_0^\dagger \mathcal{Z}(\alpha, \beta, \gamma) U'
                \big)
            \Big|^2
\end{equation}
and gradient ascent proceeds on $\{\theta_k\}, \{\varphi_k\}$ from there. Since the full simulated Hilbert space is larger than the computational space, we restrict the trace in the above to only computational states (but extend to include the $\ket{1n2}$ states in the case of the $\text{SHELVE}_{e,m}$ sequence), adjusting the normalization constant $N$ such that the maximum value attainable by $\mathcal{F}$ is 1. We additionally insert $\mathcal{Z}(\alpha, \beta, \gamma) \equiv Z_m(\gamma) Z_n(\beta) Z_e(\alpha)$ describing a set of phase rotations on each qubit--i.e. maximization of $\mathcal{F}$ results in a pulse sequence that implements $U_0$ up to single-qubit phase rotations. We list the pulse angles $\{\theta_k\}, \{\varphi_k\}$, phase correction angles $\alpha, \beta, \gamma$, and expected fidelities for chosen Rabi frequencies for each pulse sequence gate in Table \ref{tab:pulse-angles}.

\begin{table*}[t!]
    \centering
    \caption{
        \textbf{Optimized pulse sequence angles for motion-selective operations.} Full listing of the nutation ($\theta_i$), phase ($\varphi_i$), and Z-gate ($\alpha, \beta, \gamma$) angles for each gate implementation described in Appendix \ref{pulse-sequence}, with expected gate fidelity $\mathcal{F}$.
    }
    \label{tab:pulse-angles}
    \begin{tabular}{lllllllllll}
        \toprule
        Gate & $\theta_0$ & $\theta_1$ & $\theta_2$ & $\varphi_0$ & $\varphi_1$ & $\varphi$ & $\alpha$ & $\beta$ & $\gamma$ & $\mathcal{F}$ \\
        \midrule
        $\text{SWAP}_{e,m}$
            & $1.498 \pi$
            & $1.498 \pi$
            & $1.123 \pi$
            & $1.908 \pi$
            & $1.196 \pi$
            & $1.583 \pi$
            & $1.338 \pi$
            & ---
            & $1.415 \pi$
            & $\fidemswap$
            \\
        $\text{CZ}_{e,m}$
            & $1.036 \pi$
            & $1.620 \pi$
            & $1.036 \pi$
            & $0.138 \pi$
            & $0.417 \pi$
            & $0.695 \pi$
            & $1.299 \pi$
            & ---
            & $0.377 \pi$
            & $\fidemcz$
            \\
        $\text{CCZ}$
            & $1.972 \pi$
            & $1.066 \pi$
            & $0.613 \pi$
            & $0.103 \pi$
            & $0.122 \pi$
            & $0.150 \pi$
            & $1.966 \pi$
            & $1.966 \pi$
            & $1.986 \pi$
            & $\fidenmccz$
            \\
        $\text{SHELVE}_{e,m}$
            & $1.277 \pi$
            & $1.277 \pi$
            & $0.694 \pi$
            & $0.398 \pi$
            & $0.616 \pi$
            & $1.845 \pi$
            & $0.811 \pi$
            & ---
            & $0.503 \pi$
            & $\fidemshelve$
            \\
        \bottomrule
    \end{tabular}
\end{table*}

\section{Measurement protocol optimization}
\label{measurement-protocol}

We construct the quoct measurement protocol [see Fig. \ref{fig:measurement}] using a genetic algorithm to find a minimal set of operations to map computational states into the set of bit-strings describing bright/dark atomic readout results. The state readout process is inherited from previous work on the ``ququart'' space \cite{Jia2024}, where photons are scattered identically from all $\ket{0nm}$ states, but with extra considerations regarding motion. Namely, the process of scattering photons will couple to the motion degree of freedom and cause some degree of scrambling within it to any states that are bright with respect to the scattering transition. When this occurs, multiple quoct basis states can be mixed with each other and rendered indistinguishable, even through later rounds of measurement.

We account for this by using the $e$-$m$ shelving operation in Fig. \ref{fig:shelving} to swap half of the states within the subspace affected by motional scrambling to a designated set of states outside the computational space that is immune. Since no scrambling occurs along the $e$- and $n$-qubit degrees of freedom, this means the states that are not shelved are scrambled independently of all other quoct states (and each other), such that no two states are scrambled together and all quoct states remain distinguishable.


To implement genetic optimization on measurement protocols, we model a population of protocols undergoing selection and crossover where an individual protocol is a series of short circuits punctuated by a round of photon scattering as described above. We sample an initial population of 100 protocols, where individual protocols are generated by drawing a number of readout rounds from a Poisson distribution with mean 1.5, with an extra condition that the number be at least 1. For each round, a number of gates is drawn from another Poisson distribution with mean 2, and each gate is drawn uniformly from the quoct native gateset.

The initial population is evolved for 1 million generations; at each generation, the 50 least-fit protocols are discarded, and 40 new protocols are generated via crossover. Crossover between protocols is implemented as pairwise random selection between the pre-readout circuits of the parent protocols. That is, the $k$-th pre-readout circuit of the offspring is randomly selected from the $k$-th circuits of the parents with equal probability; if one parent comprises fewer circuits than the other, the remaining circuits from the other are individually either passed on or discarded. Following crossover, mutations are applied individually to all protocols whereby the number of circuits in a given protocol is increased by 1 with probability 0.3 (with the new circuit generated following the initial sampling procedure), or decreased by 1 with probability 0.3. For each circuit in the protocol, the number of gates is increased or decreased by 1, each with probability 0.3. Finally, each gate is resampled from the native gateset with probability 0.5. After mutation, 10 new protocols are generated via the initial sampling procedure to maintain population size.

Fitness is evaluated by simulating the protocol for each computational basis state and recording the resulting bitstrings of bright/dark outcomes. A composite fitness score is constructed from a boolean value $b_\t{dups}$ that is 0 if the bitstrings are unique for all basis states (1 otherwise), combined with a positive integer $f$
\begin{equation}
    f = (1 + n_\t{rounds}) (1 + n_\t{gates}) (1 + n_\t{dups})
\end{equation}
where $n_\t{rounds}$ is the number of rounds of readout, $n_\t{gates}$ the total number of gates, and $n_\t{dups}$ the number of duplicate output bitstrings. Protocols for which $b_\t{dups}$ is 0 are coarsely considered fitter than those otherwise, and $f$ is used for fine-scale comparisons within these groups.

At the end of each generation, any protocols with $b_\t{dups} = 0$ are copied to a separate population not acted upon by the algorithm. After evolution ends, all such protocols with minimal $f$ are collected. In general, multiple distinct protocols can satisfy the desired conditions, and the particular protocol presented in the main text is arbitrarily selected from this set. However, the optimal protocols our algorithm converges to share distinct characteristics, which we briefly discuss here. Namely, we find no protocols that uniquely map quoct states to bitstrings of length fewer than four. Further, we find no protocols that leave each initial computational basis state in a unique final state--that is, in addition to mapping quoct basis states to bitstrings, each protocol also constitutes a map within the quoct state space, and we find no suitable protocols for which this map is injective. Both of these characteristics may be seen as manifestations of motional scrambling. This additionally highlights the importance of the $e$-$m$ shelving operation as one that can move information out of the readout subspace--without disturbing information stored elsewhere in the computational space--such that the preserved state information can be extracted through subsequent rounds of readout. Indeed, we find no solutions when it is removed from the gate set, and no solutions that do not use it once before the first measurement.

Given a chosen measurement protocol (that shown in Fig. \ref{fig:measurement}(c), for example), analysis of the association between length-$N$ bitstrings and basis states reveals the possiblility for further optimization. Specifically, we consider the sets of input basis states to the protocol that map to the same leading substrings of length $M < N$. We find that any substring with at least one bright readout result is mapped to a set of at most two basis states and, further, that the two basis states differ by only their $n$ qubit. This is due to the nature of the shelving operation in combination with the $e$-qubit measurement as shown in Fig. \ref{fig:measurement}(a). For instance, after the initial shelving operation and assumed bright first $e$-qubit measurement in Fig. \ref{fig:measurement}(c), the quoct state is collapsed to a mixed state containing only information initially encoded in the $\ket{0n0}$ states. Given the ability to perform real-time feedback on measurement outcomes, this allows the protocol to be short-circuited by a (partial) $n$-qubit measurement as shown in Fig. \ref{fig:measurement}(b), which will measure out the remaining information in the state with just one operation. If, on the other hand, an $e$-qubit measurement is dark, no motional scrambling occurs and the protocol can proceed as shown until the next bright outcome, after which a short-circuiting $n$-qubit measurement can be applied in the same manner. Thus in some cases, full quoct readout can be performed in as few as two rounds of either $e$- or $n$-qubit readout. Below, we document the minimal bitstrings needed to identify each quoct basis state using the protocol in Fig. \ref{fig:measurement}(c), in which bright and dark outcomes yielded by an $e$-qubit readout are denoted as $B$ and $D$, while those by a short-circuiting $n$-qubit readout are $B'$ and $D'$.

\begin{equation}
    \begin{aligned}
        \ket{000} &: BB'   & \ket{010} &: BD'
        \\
        \ket{100} &: DBB'  & \ket{110} &: DBD'
        \\
        \ket{001} &: DDBB' & \ket{011} &: DDBD'
        \\
        \ket{101} &: DDDB' & \ket{111} &: DDDD'
    \end{aligned}
\end{equation}

\section{Axial gauge in the continuum to KS}
\label{axial-gauge-continuum}
Consider the locally gauge-invariant $SU(3)$ Lagrangian density \cite{Schwartz}
\[\mathcal L = -\frac1{4g^2} F^{\mu\nu a}F^a_{\mu\nu} + \bar\psi_i (i\slashed D_{ij} - m\delta_{ij})\psi_j\]
where
\[F^a_{\mu\nu} = \p_\mu A_\nu^a - \p_\nu A^a_\mu + f^{abc} A^b_\mu A_\nu^c\]
describes the color field strength indexed in the adjoint representation by $a$
and
\[D_\mu\psi_i = \p_\mu\psi_i - iA_\mu^a T_{ij}^a \psi_j\]
is the covariant derivative indexed in the fundamental representation by $i$. 
We find the canonical momenta the usual way:
\begin{align*}
\Pi^{\mu a} &= \deldel{\mathcal L}{\p_0 A^a_\mu} = \frac1{g^2}F^{\mu0 a}\\
\pi_i &= i\psi_i^\dagger.
\end{align*}
and note the primary constraint  $\Pi_0^a = 0$. This means that $A^a_0$ is non-dynamical. We can then compute the Hamiltonian density the usual way \cite{Weinberg}
\[\mathcal H = \Pi^{\mu a}\p_0 A^a_\mu + \frac1{4g^2}F_{\mu\nu}^a F^{\mu\nu a} + \pi \psi - \bar\psi(i \slashed D-m)\psi\]
We compute the gauge part first. Using
\[-g^2 \Pi^a_\mu = F_{0\mu}^a = \p_0 A_\mu^a - \p_\mu A^a_0 + f^{abc} A^b_0 A^c_\mu\]
and taking the signature $(+,-,-,-)$ we have
\begin{align*}
\p_0 A_k^a &= g^2 \Pi_k^a + \p_k A^a_0 + f^{abc} A^b_0 A^c_k
\end{align*}
and
\[\frac{1}{4g^2}F_{\mu\nu}^a F^{\mu\nu a} = -\frac{g^2}{2}\Pi_k^a\Pi^{k a} + \frac1{4g^2}F_{kl}^aF^{kla}.\]
Thus the pure gauge-field Hamiltonian density is
\begin{align*}
    &\mathcal H_A = \Pi^{\mu a}\p_0 A^a_\mu + \frac1{4g^2}F_{\mu\nu}^a F^{\mu\nu a}\\
    &= \frac{g^2}{2}\Pi^a_k\Pi^{ka} + \p_k A^a_0 \Pi^{k a} + f^{abc} \Pi^{k a}A^b_0 A^c_k + \frac1{4g^2} F^a_{kl}F^{kla}.
\end{align*}
For the matter Hamiltonian density, we obtain
\[\mathcal H_M = -i\bar\psi\gamma^k\p_k\psi - \bar\psi \slashed A^a T^a\psi + m\bar\psi\psi. \]
Now we can simplify our expression by restricting to $1+1$ dimensions and fixing an axial gauge $A_1 = 0$:
\[\mathcal H = \frac{g^2}{2} \Pi^a\Pi^a + \p_x A^a_0 \Pi^a - i\psi^\dagger\gamma^0\gamma^x\p_x\psi - A_0^a\psi^\dagger T^a\psi + m\psi^\dagger\gamma^0\psi\]
where $\Pi^a := \Pi^{1,a}$. Integrating by parts, identifying $E^a = F_{10}^a = \Pi^a$, and identifying the charged matter current $j^{\mu a} = \bar\psi \gamma^\mu T^a \psi$, we get
\[\mathcal H = -i\psi^\dagger \gamma^0\gamma^x\p_x \psi + m\psi^\dagger\gamma^0\psi + \frac{g^2}{2}(E^a)^2 -  A_0^a(\p_x E^a + j^{0 a})\]
where $A_0$ takes the form of a Lagrange multiplier enforcing Gauss' law $\p_x E^a = -\psi^\dagger T^a\psi$.

Imposing Gauss' law, taking the gamma matrix representation $\gamma^0 = \sigma_z$, $\gamma^1 = i\sigma^y$, and decomposing the spinor $\psi = (\psi_1,\psi_2)^T$, the Hamiltonian becomes
\begin{align}
H =  \int dx\,\bigg( &-i \psi_1^\dagger \p_x\psi_2 - i\psi_2^\dagger \p_x\psi_1 + m\psi_1^\dagger\psi_1 - m\psi_2^\dagger\psi_2\nonumber\\ 
& + \frac{g^2}{2}(E^a)^2\bigg)\nonumber
\end{align}
Discretizing space and staggering the fields so that $\psi_1 \rightarrow \frac1{\sqrt a}\psi_{2n}$ and $\psi_2 \rightarrow \frac1{\sqrt a}\psi_{2n+1}$, we recover the Kogut-Susskind Hamiltonian:
\[H = \sum_n\psi_n^\dagger\frac{(\psi_{n+1} - \psi_{n-1})}{ia} + m(-1)^n\psi^\dagger_n\psi_n + \frac{ag^2}{2}E_n^2.\]
We can then use the discretized Gauss' law constraint to express the chromo-electric energy in terms of the integral of the charge density in the absence of a background field:
\[E^a_{n} - E^a_{n-1} = \psi_n^\dagger T^a \psi_n =: Q^a_n.\]
With this we recover Eq.~(\ref{axial-KS}) for $\mu = 0$. An alternate derivation can be found in \cite{Sala2018}.

\section{Circuit compilation}
\subsection{Circuit identity}
\label{circuit-compilation}
Let $A$ and $B$ be two quoct operators that are hermitian, unitary, and choose a basis such that $A$ exchanges states, i.e. given the basis $\{\ket{1},...,\ket{8}\}$ there is a partition into two sets $\{\ket{i_1},...,\ket{i_4}\}$ and $\{\ket{j_1},...,\ket{j_4}\}$ such that $A\ket{i_n} = \ket{j_n}$. Then we can prove the quoct circuit identity in equation \ref{circuit-identity}:
\begin{align*}
\vcenter{\hbox{\includegraphics[width=0.8\linewidth]{circuit-identity.pdf}}}
\end{align*}
where the control is on the four states $\ket{i_n}$ and will match the transitions of $A$. Since $A$ and $B$ are hermitian and unitary, they are involutary so
\[\exp(i\alpha A\otimes B) = \cos(\alpha) I_{64} + i\sin(\alpha) A\otimes B\]
where $I_{64}$ is the $64$-dim identity matrix. Recall that
\[A\otimes B = \begin{pmatrix}
    a_{11}B & \ldots & a_{18}B\\
    \vdots & \ddots & \vdots\\
    a_{81}B & \ldots & a_{88}B
\end{pmatrix} = a_{kl}B\]
and since $A$ exchanges states, $a_{kl} = 0$ whenever $k$ and $l$ belong to the same partition. 

\begin{figure*}
    \includegraphics[width=0.8\linewidth]{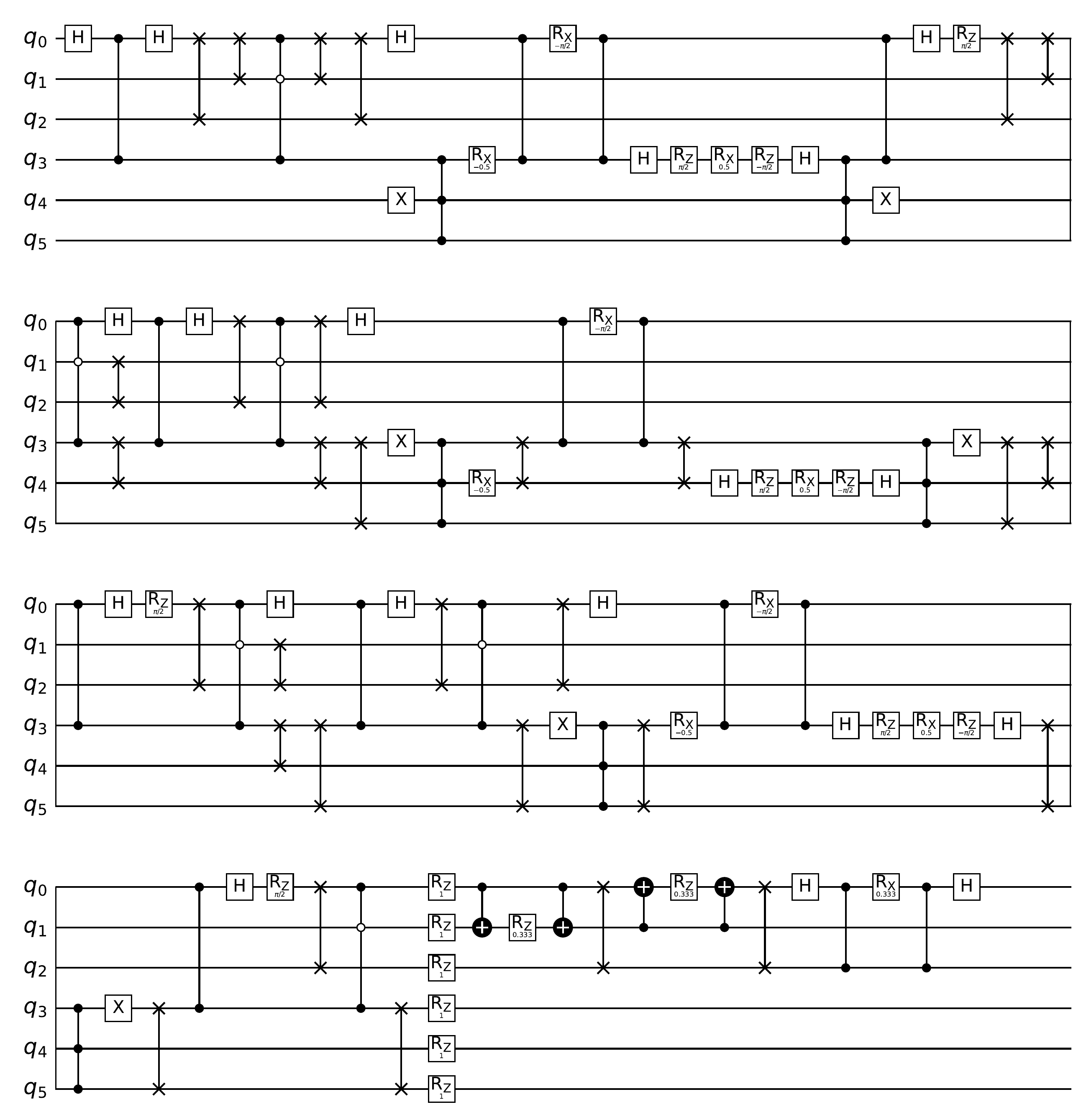}
    \caption{
        \textbf{Fully transpiled circuit for the $L = 1$ Hamiltonian}.
    }
    \label{fig:full-circuit}
\end{figure*}

Next, consider the controlled-$B$ operator $CB$. In a matrix representation $CB$ is a block diagonal matrix which has a $B$ block when $q = i_n$ and has an $I$ block when $q = j_n$. The $A(\alpha)$ gate along with the empty wire has the matrix representation 
\[\exp(i\alpha A)\otimes I_8 = \cos(\alpha)I_{64} + i\sin(\alpha) A\otimes I_8\]
Thus we want to show ${CB(A\otimes I_8)CB = A\otimes B}$. Since $CB$ is block-diagonal, left-multiplication acts on the rowspace and right-multiplication acts on the column space. Observe then that since $B^2 = I_8$,
\[CB(A\otimes I_8)CB = a_{kl}\left\{\begin{array}{cc}
    I_8 & \text{both $k,l$ in same partition}\\
    B & \text{$k,l$ in opposite partitions}
\end{array}\right.\]
since we've noted that $a_{kl} = 0$ when $k$ and $l$ are in the same partition, the equality is proven.

\subsection{Full $L=1$ circuit}
Since the circuit can be simplified by combining SWAPs and to get a sense for one Trotter step of the $L=1$ circuit, we include an example with $\mu = 0$ for completeness.

\bibliography{library}

\end{document}